%% file: main.tex
\documentclass[twocolumn,epjc3,a4paper]{svjour3}
\usepackage[utf8]{inputenc}
\usepackage{mathptmx}
\usepackage{amsmath}
\usepackage{amssymb}
\usepackage{upgreek}
\usepackage[english]{babel}

\usepackage[tight]{units}
\usepackage[dvipsnames]{xcolor}
\usepackage{graphicx}
\usepackage[colorlinks,citecolor=blue,urlcolor=blue,linkcolor=blue]{hyperref}
\usepackage[normalem]{ulem}
\usepackage{mathtools}
\usepackage{bm}
\usepackage{textgreek}
\usepackage{longtable}
\usepackage{afterpage}
\usepackage{csvsimple}
\usepackage{pgfplotstable}
\usepackage{placeins} %for the float barrier
\pgfplotsset{compat=1.9}% supress warning
\usepackage[backend=biber,style=numeric,sorting=none,url=false, minbibnames=3,maxbibnames=3,giveninits=true,isbn=false,eprint=false]{biblatex}
\AtEveryBibitem{%
  \clearfield{note}%
}

\renewbibmacro*{doi+eprint+url}{%
    \printfield{doi}%
    \newunit\newblock%
    \iftoggle{bbx:eprint}{%
        \usebibmacro{eprint}%
    }{}%
    \newunit\newblock%
    \iffieldundef{doi}{%
        \usebibmacro{url+urldate}}%
        {}%
    }

\newcommand{\cawo}{CaWO$_4$}
\newcommand{\ega}{$\upbeta$/$\upgamma$}
\newcommand{\gev}{GeV/c$^2$}

\newcommand{\erf}{\text{erf}}

\newcommand{\pars}{\vec{\theta}}

\journalname{Eur. Phys. J. C}

\input{authors_cresst_EPJ_format}
\begin{document}

%\markdownInput{Discussions/discussion_12Jan2021.md}
%\markdownInput{Discussions/discussion_19Feb2021.md}
%\clearpage

\setcounter{secnumdepth}{3} % default value for 'report' class is "2"

\title{A likelihood framework for cryogenic scintillating calorimeters used in the CRESST dark matter search}

\date{\today}
\maketitle

\begin{abstract}
  Cryogenic scintillating calorimeters are ultra\--sen\-sitive particle detectors for rare event searches, particularly for the search for dark matter and the measurement of neutrino properties. These detectors are made from scintillating target crystals generating two signals for each particle interaction. The phonon (heat) signal precisely measures the deposited energy independent of the type of interacting particle. The scintillation light signal yields particle discrimination on an event-by-event basis. This paper presents a likelihood framework modeling backgrounds and a potential dark matter signal in the two-dimensional plane spanned by phonon and scintillation light energies. We apply the framework to data from \cawo-based detectors operated in the CRESST dark matter search. For the first time, a single likelihood framework is used in CRESST to model the data and extract results on dark matter in one step by using a profile likelihood ratio test. Our framework simultaneously fits (neutron) calibration data and physics (background) data and allows combining data from multiple detectors. Although tailored to \cawo-targets and the CRESST experiment, the framework can easily be expanded to other materials and experiments using scintillating cryogenic calorimeters for dark matter search and neutrino physics.
\end{abstract}

%\tableofcontents

\section{Introduction}

Modern, precision astrophysics and cosmology hardly leave any doubt that dark matter (DM) exists and that it is five times more abundant in the Universe than ordinary, baryonic matter \cite{aghanim_planck_2020}. Unraveling the nature of DM is a major priority of particle physics. However, despite the tremendous progress made in the last decades, no unambiguous signal could be observed yet, neither at particle colliders nor through indirect or direct detection. The properties of DM, such as the mass of DM particle(s) and their interaction cross-section(s) with ordinary matter, are only loosely constrained so far. Thus, various direct DM detection experiments featuring different technologies are mandatory for a broad search for interactions of DM particles in earth-bound detectors. 

Due to their low energy thresholds, cryogenic DM detectors have leading sensitivity for light DM particles. One of them is the CRESST\footnote{\href{http://cresst-experiment.org}{cresst-experiment.org}} experiment, currently in its third stage: CRESST-III \cite{abdelhameed_first_2019}. CRESST uses scintillating crystals (typically made from \cawo) simultaneously recording a phonon (heat) signal and a scintillation light signal from a particle interaction. 

The most established and recommended tool for statistical inference, in the field of direct dark matter detection (and rare event searches in general), is the likelihood formalism \cite{baxter_recommended_2021}. This paper presents a profile maximum likelihood analysis modeling all known event types and a potential DM signal in the two-dimensional plane spanned by the phonon and light energies. We show that our likelihood framework is capable of characterizing the discrimination capabilities between the nuclear recoil signal and electromagnetic backgrounds and may also be used to combine data from different detectors. The framework presented here evolved from the one presented in \cite{angloher_results_2012}. The most crucial update is that only a single likelihood fit is done to model the data in the light-phonon plane and extract results on a potential dark matter signal, yielding proper treatment of uncertainties.

We begin with a short introduction to CRESST-II/III detectors and the data used in this work in section \ref{sec:CRESSTdetectors}, followed by an in-depth description of our likelihood framework in section \ref{sec:LikelihoodModel}. Results are presented in section \ref{sec:Results}, conclusion and outlook in section \ref{sec:Conclusion}. The presented likelihood framework is tailored to CRESST, but might, with modifications, also be applied for related experiments, in particular COSINUS\footnote{\href{http://cosinus.it}{cosinus.it}} \cite{angloher_cosinus_2016} and NUCLEUS\footnote{\href{http://nucleus-experiment.org}{nucleus-experiment.org}} \cite{strauss_gram-scale_2017}, which are based on the same technology. 

\section{CRESST low-temperature detectors} \label{sec:CRESSTdetectors}

\subsection{Detector layout and used data}

This paper uses data from CRESST-II \cite{angloher_results_2014,angloher_results_2016,reindl_exploring_2016} and CRESST-III \cite{abdelhameed_first_2019} all using \cawo~target crystals. The \cawo-crystals are equipped with a transition edge sensor (TES) made from tungsten to read the primary phonon signal from a particle interaction. We call the ensemble of target crystal and TES the phonon detector. The target crystal is paired with a cryogenic light detector made from a wafer-shaped silicon-on-sapphire piece equipped with a second, separate TES. The TES thermometers are produced at the Max-Planck Institute for Physics in Munich/Garching. 

\begin{figure}[htb]
\includegraphics[width=\columnwidth]{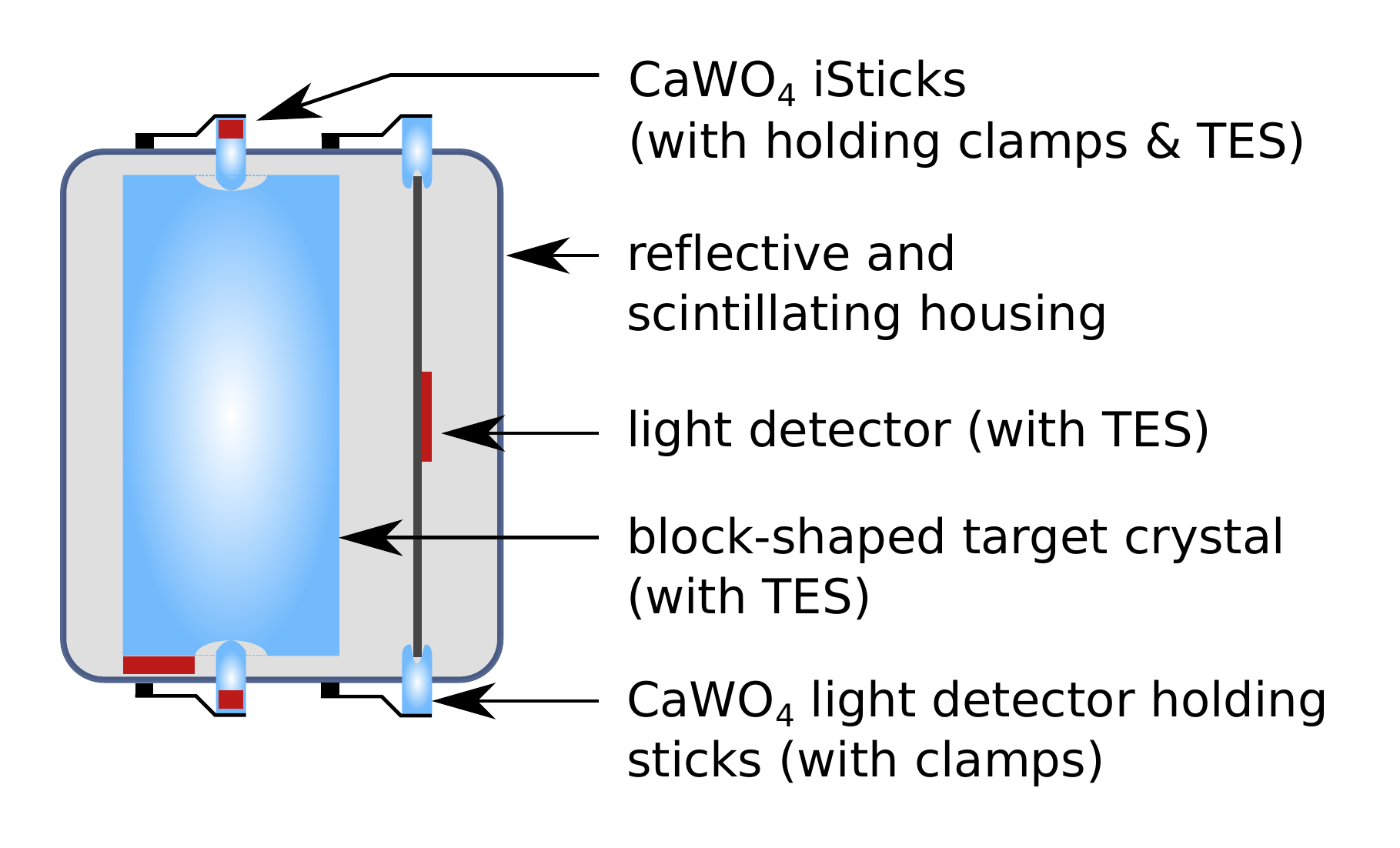}
\caption{Schematic of a CRESST-III detector module (not to scale). Parts in blue are made of \cawo; the TESs are sketched in red. The block-shaped target (absorber) crystal has a mass of \unit[$\sim$24]{g}, and its dimensions are \unit[(20x20x10)]{mm$^{3}$}. It is held by three instrumented \cawo\ holding sticks (iSticks), two at the bottom and one on top. Three non-instrumented \cawo\ holding sticks keep the square-shaped silicon-on-sapphire light detector in place. Its dimensions are \unit[(20x20x0.4)]{mm$^3$}. Figure and caption from \cite{abdelhameed_first_2019}.}
\label{fig:scheme}
\end{figure}

Phonon and light detector constitute a detector module. A schematic drawing of a CRESST-III detector module is depicted in figure \ref{fig:scheme}. Apart from the two TESs for phonon and light detector, the \cawo-holding sticks are also instrumented, thus denoted iSticks. This iStick system vetos potential background events originating from the holding. It was not yet available in CRESST-II. The iStick veto is applied during the event selection in the raw data analysis; for the likelihood framework (working on processed data) it is irrelevant. A further difference to CRESST-II are the roughly ten times lighter crystals used in CRESST-III, which were introduced to lower the energy threshold and, thus, enhance the sensitivity for light DM. Table \ref{tab:detectorsanddata} summarizes the crucial detector parameters and the data used for this work. We simultaneously fit background data (no calibration source) and neutron calibration data (AmBe neutron source) for all detectors. We use the CRESST-II detector module TUM40 to illustrate the likelihood method; afterward, we will discuss the combination of data from multiple detectors and calculate exclusion limits on the DM-nucleon cross-section using a profile likelihood ratio test.

%\onecolumngrid % the empty following line is important, enclosing in \afterpage{ } forces it to be on top of the page.

\begin{table*}[htb]
   %\begin{longtable}{ccccccccc}
   \begin{tabular}{ccrccrrcc}
    Stage      & Module name & Crystal mass   & Crystal shape & Holding scheme             & Threshold       & Exposure & Producer & Ref. \\
    \hline\\
    CRESST-II  & TUM40            & \unit[249]{g}  & Block-shaped  & \cawo~sticks               & \unit[$(605.6\pm4.4)$]{eV}  & \unit[98.0]{kgd}     & TUM     & \cite{reindl_exploring_2016} \\
    CRESST-II  & Lise               & \unit[300]{g}  & Cylindrical   & Bronze clamps              & \unit[$(307.3\pm3.6)$]{eV}  & \unit[52.2]{kgd}     & commercial     & \cite{angloher_results_2016} \\
    CRESST-III & Detector A         & \unit[23.6]{g} & Block-shaped  & \cawo~iSticks  & \unit[$(30.1\pm0.1)$]{eV} & \unit[5.89]{kgd}     & TUM  & \cite{abdelhameed_first_2019} \\
  %\end{longtable}
  \end{tabular}
  \caption{Overview on the detector modules used in this work. All target crystals were made from \cawo. TUM = Technical University Munich}
  \label{tab:detectorsanddata}
  \end{table*} 

%\twocolumngrid

\subsection{Energy, light and light yield information}

Before we present the construction of the likelihood framework, we introduce the relevant observables. For every particle interaction, a CRESST detector simultaneously measures the energies deposited in the phonon channel ($E_\text{p}$) and in the light channel ($L$), respectively. We define the light yield $LY$ as the ratio of these two quantities: $LY=L/E_\text{p}$, which is set to one for events from a predetermined calibration line\footnote{\unit[122]{keV} $\upgamma$s from $^{57}$Co for CRESST-II and \unit[63.2]{keV} $\upgamma$s from the K$_\upalpha$-escape of tungsten for CRESST-III when illuminated with $^{57}$Co.}. This calibration results in an electron-equivalent unit, keVee, for $E_\text{p}$ and $L$. However, other event types are characterized by different light yields. In particular, the so\-ught\--for nuclear recoils exhibit lower light yields than the dominant \ega-backgrounds. Since we measure the energy sharing between phonon energy $E_\text{p}$ and light energy $L$, we can obtain the total event-type-independent energy, in keV, using the following equation: 
\begin{equation}
    E = \eta L + (1-\eta) E_\text{p}  = [1-\eta(1-LY)]E_\text{p} 
    \label{eq:phononantiquenching}
\end{equation}
The scintillation efficiency $\eta$ quantifies the amount of energy converted to scintillation light for a $\upgamma$-calibration event. As one can see, the above formula will affect all events with a light yield not equal to one and allows to correctly obtain the total deposited energy $E$ for any event in the detector  \cite{angloher_results_2014,reindl_exploring_2016}. In this work, the scintillation efficiency $\eta$ is a free parameter in the maximum likelihood fit. The obtained values are compatible with previous results \cite{angloher_results_2014,reindl_exploring_2016} where $\eta$ was extracted from the tilt of $\upgamma$-lines in the light-phonon energy plane caused by statistical fluctuation of the light production. 

Throughout the paper, we will stick to the following notation:
\begin{description}
         \item[$L$:] Energy measured in the light detector (keVee) 
         \item[$E_\text{p}$:] Energy measured in the phonon detector (keVee)
         \item[$E$:] Event-type-independent total deposited energy (keV)
         \item[$LY$:]  Light Yield $LY=L/E_\text{p}$ 
\end{description}

\section{Likelihood framework} \label{sec:LikelihoodModel}

We use the extended maximum likelihood framework (explained in more detail in subsection~\ref{subsec:PRL}) to model backgrounds and a potential signal in the $L$-$E$-plane using non-normalized density functions $\rho_x(E,L,\pars_x)$ (with $x$ standing for different density functions). They depend on $E$, $L$ and the parameters $\pars_x$. Most density functions, except for the excess light events (see subsection \ref{subsec:excesslight}), can be written as the product of a band describing the position in the $L$-$E$-plane and an energy spectrum ($dN/dE_x(\pars_x)$) accounting for the content of the band. A band is given by its mean line ($L_x(E,\pars_x)$) (subsection \ref{subsec:bandmean}) and a Gaussian function with energy-dependent $\sigma_x(E)$ to describe its width (subsection \ref{subsec:bandwidths}). 

\begin{figure}[h!]
    \centering
    \includegraphics[width=\columnwidth]{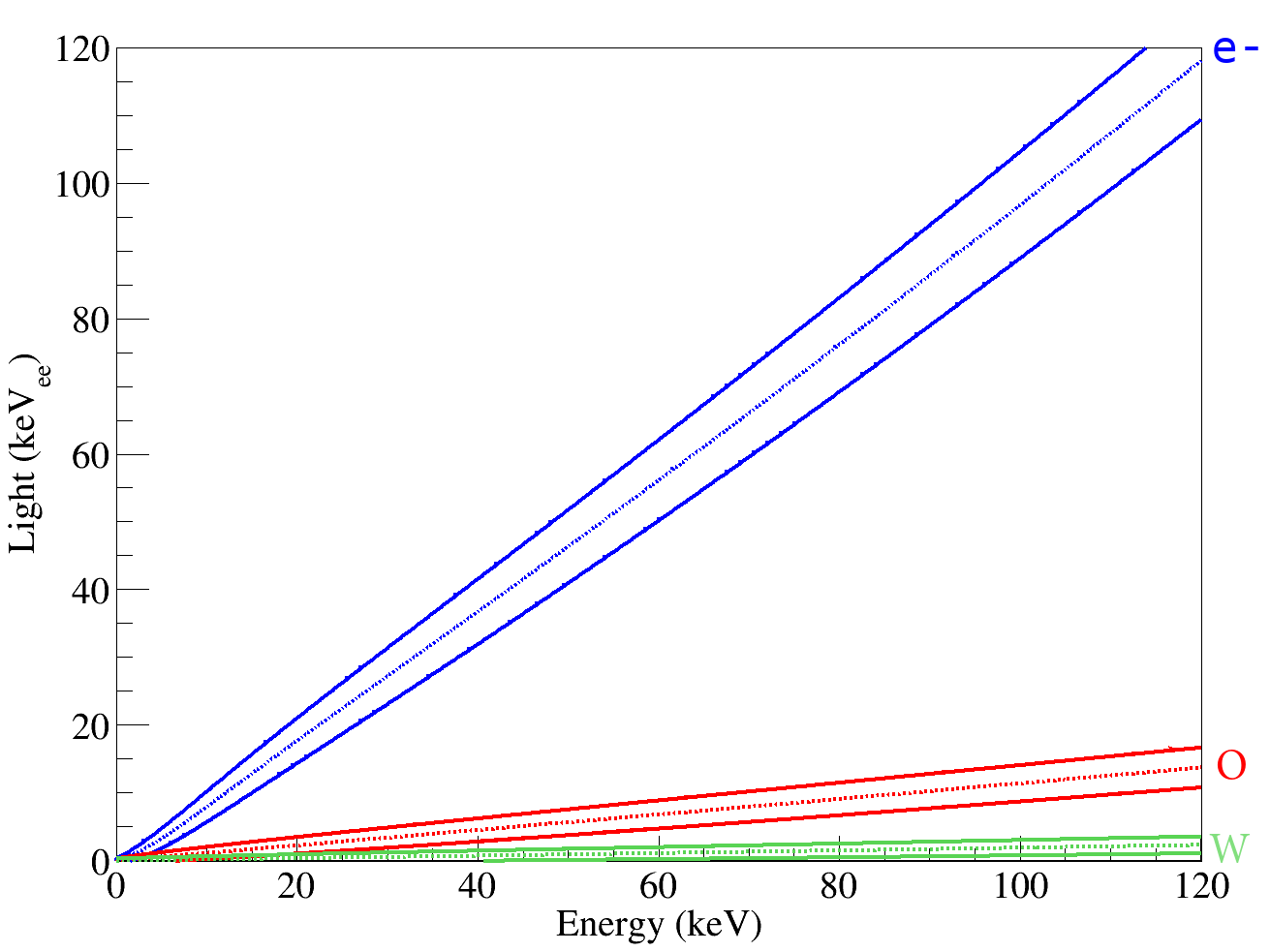}
    \caption{Schematic illustration of the electron band (blue) and the nuclear recoil bands for oxygen (red) and tungsten (green) in the light ($L$) versus energy ($E$) plane. The calcium band is not shown for clarity; it would be located between the oxygen and the tungsten band.}
    \label{fig:illustrationLight}
\end{figure}

\begin{figure}[h!]
    \centering
    \includegraphics[width=\columnwidth]{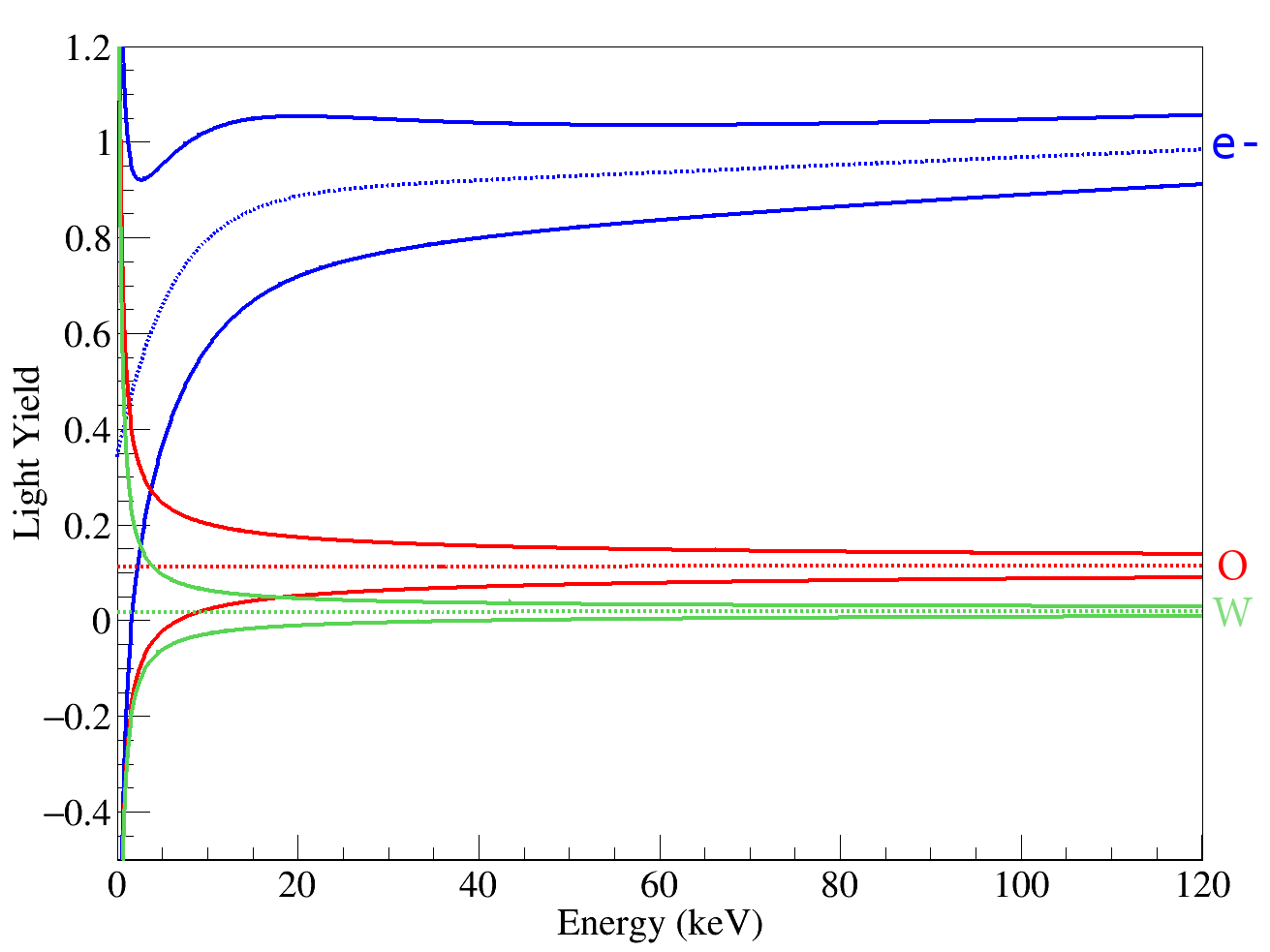}
    \caption{Schematic illustration of the same bands (electron band (blue), oxygen band (red) and tungsten band (blue)) as in figure \ref{fig:illustrationLight} but in the light yield ($LY$) versus energy ($E$) plane.  The calcium band is not shown for clarity; it would be located between the oxygen and the tungsten band.}
    \label{fig:illustrationYield}
\end{figure}

For illustration, we show exemplary bands in the $L$-$E$-plane in figure \ref{fig:illustrationLight} and the same bands in the $LY$-$E$-plane in figure \ref{fig:illustrationYield}. Both figures show the mean lines (dashed) and the upper and lower \unit[90]{\%} boundaries (solid) which are the \unit[$\pm$1.28]{$\sigma$} one-sided quantiles around the (dotted) mean line. Within the two solid boundaries \unit[80]{\%} of the events of the respective event class are expected (with respect to their distribution in $L / LY$ for a given energy $E$). %The energy spectra (contents of the bands) will be presented in subsection \ref{subsec:energyspectra}.


\subsection{Mean of the bands} \label{subsec:bandmean}

In this subsection we describe the mean of the bands in the $L$-$E$-plane for each of the respective event classes. The width of the bands will be discussed in the following subsection \ref{subsec:bandwidths} and the corresponding energy spectra (the content of the bands) in subsection \ref{subsec:energyspectra}.

\subsubsection{Electron band}
All bands are derived from the electron band (or $\upbeta$-band) which describes energy depositions in the crystal stemming from interactions with electrons in the crystal. The electron band is the band with the highest light yield and its mean line $L_\text{e}(E)$ is modeled as:  
\begin{equation}\label{eq:eband}
    L_\text{e}(E)=\left(L_0E+L_1E^2\right)\left[1-L_2\exp\left(-\frac{E}{L_3}\right)\right]
\end{equation} 
$L_0$ accounts for the linear light output while $L_1$ is an empirical parameter allowing small deviations from linearity. The term in square brackets is called non-proportionality arising from the saturation of light production per unit path length %$dL/dx$
at low energies, see \cite{lang_scintillator_2009,schmaler_cresst_2010} for details. The non-proportionality effect varies for different \cawo-crystals and is typically more pronounced for crystals grown at TU Munich (TUM40 and Detector A) than for commercially produced ones (Lise).

\subsubsection{$\upgamma$-band} \label{subsubsec:gammaband}

$\upgamma$s interacting in the target crystal transfer their energy to multiple electrons. For the energy $E$ this makes no difference, as only the total deposited energy matters. However, because of the aforementioned non-proportionality effect, the total amount of produced light $L$ is smaller if multiple electrons share the energy. We found, using data and a Monte Carlo simulation, that the mean line of the $\upgamma$-band $L_{\upgamma}(E)$ can be modeled by evaluating the electron band at a reduced energy
\begin{equation}\label{eq:gbandexp}
    L_{\upgamma}(E)=L_\text{e}\left(E\left[Q_{\upgamma,1}+E \, Q_{\upgamma,2}\right]\right)
\end{equation} 
with two detector-specific free fit parameters $Q_{\upgamma,1/2}$.

\subsubsection{$\upbeta$/$\upgamma$ band(s)}

We also observe mixtures between $\upbeta$ and $\upgamma$ events in the case of a $\upbeta$-decay accompanied by the emission of a $\upgamma$ from the subsequent de-excitation of the nucleus.\footnote{The time resolution of cryogenic detectors is by far not sufficient to resolve the energy depositions of the $\upbeta$ and the $\upgamma$; their energies sum up in a single event.} The total deposited energy of such a mixed event is $E=E_\upgamma + E_\upbeta$ and the total light consequently is:
\begin{eqnarray} \label{eq:betagammaquenching}
       L_{\upbeta/\upgamma,i}(E) = L_\upgamma(E_{\upgamma,i}) + L_\text{e}(E-E_{\upgamma,i})\\
       i: \text{index of $\upbeta/\upgamma$-decay in respective detector}  \nonumber 
\end{eqnarray}
This description is specific for each excited state of each $\upbeta$-decaying nuclide, thus we added the index $i$.

\subsubsection{Elastic nuclear recoil band(s)}
Nuclear recoils exhibit a significantly lower light output than electron events, an effect known as quenching. The quenching depends on the nuclide resulting in one band per nuclide in the target material. The reduction in light output of a nuclear recoil compared to an electron recoil of the same energy is quantified by the quenching factor which may depend on the energy. Nuclear recoils do not suffer from the saturation effects%in $dL/dx$
, thus they are not affected by the non-proportionality effect. The mean lines of the elastic nuclear recoil (ENR) bands are then obtained by: 

\begin{eqnarray}\label{eq:meann}
    L_{\text{ENR},x}(E)=\left(L_0E+L_1E^2\right) \cdot \epsilon \cdot QF_{x} \nonumber \\ \cdot  \left[1+f_x\exp\left(-\frac{E}{\lambda_{x}}\right)\right]\\ x \in {\text{O,Ca,W}} \nonumber
\end{eqnarray}

In the above equation, $\left(L_0E+L_1E^2\right)$ describes the electron band without non-proportionality, and 
\begin{equation*} QF_{x} \cdot \left[1+f_x\exp\left(-\frac{E}{\lambda_{x}}\right)\right] \end{equation*}
parametrizes the quenching factors measured for the three nuclei in \cawo~with a neutron beam, see \cite{strauss_energy-dependent_2014} for the measurement and details on the parametrization. The corresponding parameter values are listed in table \ref{tab:QFParameters}. Since the neutron beam measurement could determine these parameters with high precision, they are fixed in our likelihood fit. However, in \cite{strauss_energy-dependent_2014} the authors also found that different \cawo-crystals may have shifted nuclear recoil bands, where all three bands are commonly affected. Thus we add the detector-specific free fit parameter $\epsilon$ to our model. We obtain values for $\epsilon$ that are compatible with \cite{strauss_energy-dependent_2014}.
    
%\subsubsection{Quenching of nuclear recoil bands}
    \begin{table}[!h]\
    \centering
    \begin{tabular}{cccc}
    $x$                      & $QF_x$  & $f_x$   & $\lambda_x$ \unit[]{(keV)} \\ \hline
    O   & $0.07391\pm0.00002$ & $0.7088\pm0.0008$ & $567.1\pm0.9$                    \\ 
    Ca  & $0.05560\pm0.00073$ & $0.1887\pm0.0022$ & $801.3\pm18.8$                    \\ 
    W   & $0.0196\pm0.0022$ & $0$      & $\infty$                       \\ 
    \end{tabular}
    \caption{Parameters for the quenching of nuclear recoils in \cawo~as measured in \cite{strauss_energy-dependent_2014}. Please note that values in \cite{strauss_energy-dependent_2014} were parametrized in $LY_x$ which we converted to $QF_x$ here.}
    \label{tab:QFParameters}
\end{table}

\subsubsection{Inelastic nuclear recoil band(s)}
We find inelastic nuclear recoils (INRs) in the neutron calibration data. These are interactions of neutrons with atomic nuclei where the nucleus is left in an excited state and de-excites with the emission of a $\upgamma$-ray or an electron. Since de-excitation happens quickly, our detectors will measure the total deposited energy in the crystal which is the sum of the kinetic energy transferred to the nucleus plus the energy released by the de-excitation. The light output of an INR is thus a mixture of the light output of a nuclear recoil (off the respective nucleus) and the output of a $\upgamma$ or electron. Consequently, the INRs are modeled analogous to the $\upbeta/\upgamma$-events (see equation \ref{eq:betagammaquenching}): 

\begin{eqnarray} \label{eq:inelasticsquenching}
   L_{\text{INR},i}(E) = L_{z}(E_{z,i}) + L_{\text{ENR},x}(E-E_{z,i})\\
   i: \text{index of INR band in respective detector} \nonumber \\
   z \in {\upbeta,\upgamma}:  \text{$\upbeta$ (electron) or $\upgamma$ from de-excitation} \nonumber \\ 
   x \in {\text{O,Ca,W}} \nonumber
\end{eqnarray}

\subsubsection{Constant light yield or no-light band}

Although trivial, we explicitly mention that a band with a constant $LY$ ($LY_\text{c}$) can be defined: 

\begin{equation} \label{eq:nolightband}
    L_{LY_\text{c}} = LY_\text{c} \cdot E
\end{equation}

$LY_\text{c}=0$ corresponds to a no-light band for events where no scintillation light is produced.

\subsection{Width of the bands} \label{subsec:bandwidths}
The distribution of the events in the band at a specific energy (width of each band) is modeled as a Gaussian function around its mean line with an energy-dependent width. It results from the finite resolutions of light and phonon detector, where the impact of the latter depends on the slope of the band in the $L$-$E$-plane (equation \ref{eq:sigmacomplete}). 
\subsubsection{Resolution of the light detector}
The resolution of the light detector $\sigma_L(L)$ is given by:
%\subsubsection{Resolution of the light detector}
\begin{equation}\label{eq:sigl}
\sigma_\text{L}\left(L\right)=\sqrt{\sigma_\text{L,0}^2 + S_1 L + S_2 L^2}
\end{equation} 
In the above equation $\sigma_{L,0}$ is the baseline resolution (= resolution at zero light, $\sigma_L(0)$) of the light detector. This parameter is determined precisely from the raw data analyses \cite{abdelhameed_first_2019,angloher_results_2016,reindl_exploring_2016} and, therefore, fixed in the likelihood fit. Instead, $S_1$ and $S_2$ are free fit parameters. $S_1L$ results from the Poissonian light production and detection process and consequently enters $\sigma_L$ proportional to $\sqrt{L}$.  $S_1$ is the amount of energy that has to be deposited in the target crystal to measure one photon in the light detector (on average and for an electron event). $S_2$ is a generic parameter that was empirically found to improve the fit result for certain detectors. However, $S_2$ is typically small, and the total resolution $\sigma_{L}(L)$ is dominated by $\sigma_{L,0}$ close to the threshold and by $S_1L$ otherwise.\footnote{This means that the width of the bands can be approximated by a Gaussian function for the whole energy range, as long as the light detector cannot resolve individual photons. %Since a Gaussian has a less pronounced tail than a Poissonian distribution, we may slightly underestimate the leakage of the \ega-band to the nuclear recoil bands, which is conservative for limit setting.
The Gaussian approximation is used to reduce complexity and to keep the computational costs at an acceptable level.} 

\subsubsection{Resolution of the phonon detector}
The resolution of the phonon detector $\sigma_P(E)$ is parametrized in the following way: 
%\daniel{ich bin immer noch der meinung das wir die andere resolution parameterisierung verwenden sollten, schon alleine wegen der convolution um conservative zu sein}
\begin{equation}\label{eq:sigp}
   % old formula 
    \sigma_\text{P}\left(E\right)=\sqrt{\sigma_\text{P,0}^2 + \sigma_\text{P,1}^2 \left(E^2 - E_\text{thr}^2\right)}
   % \sigma_P\left(E\right)=\sqrt{\sigma_{P,0}^2 + (\sigma_{P,1}E)^2 }
\end{equation}
It is dominated by the baseline resolution $\sigma_{\text{P},0}$ which again is determined externally and fixed in the fit, while $\sigma_{\text{P},1}$ remains free. $\sigma_{\text{P},1}$ is typically small and accounts for a potential energy dependence of the resolution. $E_\text{thr}$ is the energy of the trigger threshold.

\subsubsection{Total width of the bands}
The width of each band is then calculated by: 
\begin{eqnarray} \label{eq:sigmacomplete}
  \sigma_x\left(E\right)=\sqrt{\sigma_\text{L}^2\left(L_x\left(E\right)\right) + \left(\frac{\text{d}L_{x}}{\text{d}E}\left(E\right)\right)^2\sigma_\text{P}^2\left(E\right)}\\
%x \in ENR,\text{O,Ca,W,}; INR,{\text{O,Ca,W,}; \gamma; \gamma/\beta; e} \nonumber
  \text{$L_x$ may be any band, see equations \ref{eq:eband}, \ref{eq:gbandexp}, \ref{eq:betagammaquenching}, \ref{eq:meann}, \ref{eq:inelasticsquenching}, \ref{eq:nolightband}. } \nonumber
\end{eqnarray}
The slopes $\text{d}L_{x}/\text{d}E$ are calculated analytically for maximum speed and accuracy. Figure \ref{fig:illustrationLight} intuitively illustrates that steeper bands (higher slope $\frac{\text{d}L_{x}}{\text{d}E}$) are more affected by the resolution of the phonon detector ($\sigma_\text{P}(E)$).

\subsection{Energy spectra} \label{subsec:energyspectra}
While the two previous subsections (\ref{subsec:bandwidths} and \ref{subsec:bandmean}) described the position of the various bands, this subsection will present the modeling of their content and their energy spectra. For this work, we are using an empirical model. However, for future work, we may also include results from Geant4 Monte Carlo simulations used for the CRESST background model  \cite{abdelhameed_geant4-based_2019,angloher_high-dimensional_2023}.

\subsubsection{Electron spectrum}

Electrons from Compton scattering dominate the electron band. Their energy spectrum for the energy regime of interest is mostly constant with a small linear slope and consequently given by:
    \begin{equation}\label{eq:espece}
      \frac{\text{d}N_\text{e}}{\text{d}E}=P_0+E P_1
    \end{equation}
    
\subsubsection{$\upgamma$-peaks}
We find several $\upgamma$-peaks in our physics data. Each of them is modeled with a Gaussian function of free mean $\mu$ and amplitude $C$. Their width is given by the resolution of the phonon detector $\sigma_\text{P}(E)$ (see equation \ref{eq:sigp}). In total we get the following equation: 
\begin{eqnarray} \label{eq:gammapeak}
  \frac{\text{d}N_{\upgamma,i}}{\text{d}E}= \frac{C_i}{\sqrt{2\pi}\sigma_\text{P}\left(\mu_{i}\right)}\exp\left(-\frac{\left(E-\mu_{i}\right)^2}{2\sigma^2_\text{P}\left(\mu_{i}\right)}\right) \\
i: \text{index of $\upgamma$-peak in respective detector}  \nonumber
\end{eqnarray}

The determination of the $\upgamma$-peaks to be included in our model is an iterative process. We start by fitting a likelihood framework including all known peaks to the data, simultaneously allowing for determining the position of the $\upgamma$-band. All events located inside of the \unit[90]{\%} upper and lower boundary of this band are then selected, and a density function is created either through binning or Gaussian kernel density estimation (KDE). The width of the Gaussian kernels, as well as the width of the bins, is set according to the resolution of the phonon detector. We find that the Gaussian KDE method usually yields better results. In comparison, the Gaussian KDE had a higher ratio of known peaks to overall found peaks. Therefore, we use it for all the data presented in this paper. Peaks are then localized in that spectrum based on their topographic prominence \cite{_topographic_2022,tungli_findpeaks.jl_2022}. The possible origins of additional, a priori unknown peaks are then determined by comparing their energy to known $\upgamma$-sources. We include any peak of plausible origin in the final model.

%\daniel{der algorithmus um die peaks aus dem spektrum zu finden ist dieses:
%https://github.com/tungli/Findpeaks.jl
%keine ahnung ob und wie wir das am besten referenzieren}
%Plots: Im Artikel: Fehler bei Achsenbeschriftung LY = L/Ep, Scintillation Light YIELD
%          Konsistent "light yield" verwenden, nicht "lightyield"
%       - TUM40: LY vs. E, separat fuer bck und ncal
%       - TUM40: Espec, separat fuer bck und ncal
%       - TUM40: LY hist, separat fuer bck und ncal
%Plots: Appendix
%.      - Wie oben fuer Lise und Det A, aber LY vs. E kombiniert 
%       - TUM40 bis 400keV
% Fit results for other detectors as ancillary csv files 

\subsubsection{$\upbeta$/$\upgamma$ spectra}
The $\upbeta/\upgamma$ spectra consist of a $\upgamma$-photon with energy $E_{\upgamma,i}$ and a decaying tail down to the Q-value of the respective $\upbeta$-decay. Therefore, we model the $\upbeta/\upgamma$-spectra as a triangle starting at $E_{\upgamma,i}$ and approaching zero for the Q-value. To account for the resolution this triangle is convolved with $\sigma_\text{P}(E)$ (equation~\ref{eq:sigp}): 

%\flo{ @Daniel: Ich kenn mich hier nicht aus welcher Parameter was ist. z, QmEz, Bitte hinschreiben. Meine Formel steht unten auskommentiert, aber die ist ja leicht anders was die Amplitude angeht. }
%\daniel{hoffe es ist jetzt verständlich aber ich habe absolut keine Ahnunung wieso es da errors bei der Formel gibt, habe die Klammern mehrmals überprüft}
%\flo{Da waren zwei Fehler. Einmal ein left ohne Klammer und die eckige Klammer ging über mehrere Zeilen was aber glaube ich formal mehrere equations sind.}
%\daniel{left ohne klammer war nötig um die große eckige klammer zu anzuzeigen}

\begin{eqnarray}  \label{eq:betapeak}
  \frac{\text{d}N_{\upbeta/\upgamma,i}}{\text{d}E}= \frac{C_i}{D_i}  \bigg[ \left(1-\frac{E'_i}{D_i}\right)  \\
    \cdot \left[\erf\left(\left(D_i - E'_i\right)  z\right) - \erf\left(-E'_i  z\right)\right]  \nonumber \\
    + \frac{\exp\left( \left[-\left(D_i-E'_i\right)z\right]^2 \right)-\exp\left(\left[E'_i z\right]^2\right)}{z  \sqrt{\pi}  D_i}\bigg]  \nonumber \\
    \text{with:} \\ \nonumber
    E'_i = E - E_{\upgamma,i} \\ \nonumber
    z = \frac{1}{\sigma_\text{P}(E)  \sqrt{2}} \\  \nonumber
    D_i = Q_i - E_{\upgamma,i} \\ \nonumber
    i: \text{index of $\upbeta/\upgamma$-spectrum in respective detector}  \nonumber
    %\frac{dN_{\beta,x}}{dE}= \frac{C_x}{D_x} * \left[ \left(1-\frac{E}{D_x}\right)  \\
    %* \left[erf\left(\left(D_x - E\right) * z\right) - erf\left(-E * z\right)\right]  \nonumber \\
    %\left + \frac{exp\left( \left[-\left(D_x-E\right)*z\right]^2 \right)-exp\left(\left[E*z\right]^2\right)}{z * \sqrt{\pi} * D_x}\right]  \nonumber \\
    %z = \frac{1}{\sigma_p * \sqrt{2}}  \nonumber
   % C_i / (E_{\gamma,i} - Q_i) * ((E - Q_i) / 2. * (erf( \frac{E - E_{\gamma,i}}{\sqrt{2} \sigma_P(E)}) - \\ 
%    erf( \frac{E-Q}{\sqrt{2} \sigma_P(E)}})) - \frac{\sigma_P(E)}{\sqrt{2}\pi} * \\
 %   (\exp{( \frac{(E-Q_i)^2}{2 \sigma_P(E)^2}) - exp(\frac{(E-E_{\gamma,i})^2}{\sigma_P(E)^2})))
\end{eqnarray}

Typically the pre-factor $C_i$ and the energy of the $\upgamma$ particle $E_{\upgamma,i}$ are free fit parameters, while the distance between $\upgamma$ and Q-value $D_i = Q_i - E_{\upgamma,i}$ is fixed in the fit to the literature value. This parametrization allows scaling the amplitude and shifting the convolved triangle but not shrinking or stretching it.

\subsubsection{Bremsstrahlung spectrum}

$\upgamma$s are also created as secondary particles from high-energy charged particles, particulary muons, interacting in the copper surrounding the detectors. Our understanding is that the high-energy particles create $\delta$-electrons that emit Brems\-strahl\-ung when slowed down \cite{akkerman_delta-electron_2014}. The result is a characteristic bump peaking at about \unit[150]{keV} \cite{heusser_low-radioactivity_1995}. Although the CRESST setup is equipped with a muon veto, a faint bump may still be observable in the background data. The neutron data, on the other hand, show a clear bump which is confirmed by Geant4 Monte Carlo simulations \cite{fuss2022}.\footnote{The main contribution are secondary $\upgamma$s generated during the illumination with the AmBe neutron source. In addition, for technical reasons, the lead/copper shielding may not be fully closed during AmBe calibration, and also the muon veto is not in use during the neutron calibration.} To model the bump we use a semi-empirical description of an exponential decay multiplied with a third-order polynomial:

\begin{eqnarray}
  \frac{\text{d}N_\text{gb}}{\text{d}E} = A_\text{gb} \cdot \exp(p_\text{gb,exp} E)   \\  \cdot ( 1 + p_\text{gb,1} E + p_\text{gb,2} E^2 + p_\text{gb,3} E^3) \nonumber
\end{eqnarray}

We use the same shape parametrization for background and neutron calibration data (parameter values $p_{\text{gb},x}$). The amplitudes $A_{\text{gb}}$ differ for background and neutron data to account for the much higher rate of these events during the neutron calibration.

\subsubsection{Elastic nuclear recoil spectra}

In the neutron calibration data, we model the energy distributions of neutron-induced ENRs as exponential functions 
\begin{eqnarray}
  \frac{\text{d}N_{\text{ENR},x}}{\text{d}E} = A_x \cdot \exp(-E / l_x) \\
    x \in {\text{O,Ca,W}} \nonumber
\end{eqnarray}
with three fit parameters $A_x$ and $l_x$. For the background data, the neutron background is very small and thus set to zero in the likelihood fit.

\subsubsection{Inelastic nuclear recoil spectra}

The energy spectrum of INRs is modeled as a linear function with the constant $P_{0,\text{INR}}$ and the parameter $P_{1,\text{INR}}$ varying linearly with the distance to the excitation energy $S_\text{INR}$. To account for the resolution, we convolve this function with the resolutions at $S_\text{INR}$ and the Q-value (denoted $E_\text{INR}$ here), resulting in the error functions in equation \ref{eq:ielastics}.

 \begin{eqnarray}  \label{eq:ielastics}
   \frac{\text{d}N_{\text{INR},i}}{\text{d}E} = [P_{0,\text{INR},i} + (E-S_{\text{INR},i}) P_{1,\text{INR},i}] \\ 
    \cdot \frac{1}{2} \left[ \erf\left(\frac{E-S_{\text{INR},i}}{\sqrt{2} \sigma_\text{P}(S_{\text{INR},i})}\right) - \erf\left(\frac{E-E_{\text{INR},i}}{\sqrt{2} \sigma_\text{P}(E_{\text{INR},i})}\right) \right] \nonumber
    \end{eqnarray}
In the above equation, we neglect subdominant terms to enhance computing speed at practically no loss of accuracy. Like for ENR, the number of INR events in background data is negligible and, consequently, set to zero in the likelihood fit.
    
\subsubsection{Low-energy-excess spectrum} \label{subsubsec:lee}
We observe an exponentially rising event population close to the threshold of some detectors, typically denoted low-energy-excess (LEE) \cite{adari_excess_2022,angloher_latest_2023}. The origin of these events is not fully understood so far, and they appear at very low energies where hardly any scintillation light is produced. We model the LEE with a no-light or constant light yield band (equation \ref{eq:nolightband}) and (one or two) exponential functions for their energy spectrum:

\begin{equation}\label{eq:ecxessspectrum}
  \frac{\text{d}N_\text{lee}}{\text{d}E}=\sum_i^{N_i} F_{\text{e},i}\exp\left(-\frac{E}{l_{\text{e},i}}\right)
\end{equation}
$N_i$ is the number of exponentials used to model the LEE.

Under standard assumptions, the expected DM-nucleus recoil spectrum features an approximately exponential shape. Therefore, any potential background with an exponentially shaped energy spectrum must be treated with special caution since it can be misinterpreted as a signal and vice versa. While a discussion of the LEE is not the goal of this work, there are strong indications for a non-DM origin of the LEE in CRESST-III \cite{angloher_latest_2023}. Also, many other rare event searches see a similar excess which led to a community effort in the form of a workshop series denoted "EXCESS" \cite{adari_excess_2022,_excess_2021,_excess2022_2022,_excess22idm_2022}. Furthermore, due to the large abundance of LEE events in many detectors, a description of the data in a maximum likelihood framework is not reasonably possible without accounting for this population.

We find that the LEE in TUM40 is reasonably modeled with one exponential, while Detector A benefits from two exponentials to describe the LEE. We show the effect of the LEE and its modeling on the DM exclusion limits in the appendix \ref{sec:lee_effect}.

It should be noted that in the case of the profile likelihood calculation of an exclusion limit where the signal contribution is gradually increased from the best-fit point (while the rest of the parameters in the model are kept free) until the model is no longer compatible with the data, the contribution assigned by the likelihood to the LEE gradually decreases if there is any ambiguity between DM signal and LEE background.

\subsubsection{Potential dark matter signal}

We follow the so-called standard scenario to model a potential DM signal, assuming elastic, coherent DM-nucleus scattering and an isothermal, Maxwellian dark matter halo. It is given by

\begin{eqnarray} \label{eq:dmsignal}
  \frac{\text{d}N_{\upchi,x}}{\text{d}E} = \frac{\rho_\upchi}{2 m_\upchi \mu^2_\text{N}} \sigma_0 F^2(E) \int_{v_{\text{min}(E)}}^{v_\text{esc}} d^3v \frac{f(\vec{v})}{v} \\ x \in {\text{O,Ca,W}} \nonumber
\end{eqnarray}
with the DM-nucleus scattering cross-section $\sigma_0$, the DM mass $m_\upchi$, the reduced DM-nucleus mass $\mu_x$, the local DM density $\rho_\upchi$, and the form factor $F(E)$. To be able to compare results of different experiments, not the DM-nucleus cross-section is reported, but the DM-nucleon $\sigma_\upchi$ cross-section:

\begin{eqnarray}
  \sigma_\upchi = \left(\frac{1 + m_\upchi/m_x}{1 + m_\upchi/m_\text{p}}\right)^2 \frac{\sigma_0}{A^2}
\end{eqnarray}
where $m_x$ is the mass of the nucleus and $m_\text{p}$ is the proton mass.
The integral over the velocity distribution $f(\vec{v})$ starts at the minimal velocity to produce a nuclear recoil above $E$, $v_\text{min}(E) = \sqrt{ (E m_x) / (2 \mu_\text{N}^2)}$ and runs to the galactic escape velocity $v_\text{esc}$. For $f(\vec{v})$ we follow the implementation in \cite{donato_effects_1998}, with $v_\text{esc}=\unit[544]{km/s}$, a(n) (azimuthal) solar velocity of $v_\text{sun}=\unit[231]{km/s}$, a velocity at infinite distance  $v_\text{inf}=\unit[220]{km/s}$ and the dark matter density at Earth's position $\rho_\upchi=\unit[0.3]{(GeV/c^2)/cm^3}$. We average over the annual modulation effect caused by the earth's orbit around the sun.\footnote{Including time information for signal and background is subject of future work and beyond the scope of this article}%, which is justified given the typical length of CRESST measurements being one year or more. 
This implementation of the expected DM signal is identical to the one previously used by CRESST \cite{abdelhameed_first_2019, angloher_results_2014,angloher_results_2016} and follows to a large extent the recommendation of PHYSTAT-DM \cite{baxter_recommended_2021}. As recommended by PHYSTAT-DM, we do not profile over the uncertainties of the DM halo, although technically possible.

For high momentum transfer, coherence is lost, and the form factor $F(E)$ of the nucleus is considered. For oxygen and calcium, we use the model-independent para\-metri\-zation of the form factors presented in \cite{duda_model-independent_2007}. For tungsten, a Woods-Saxon approximation is made whose parameters may also be found in \cite{duda_model-independent_2007}.\footnote{Details on the form factors for \cawo~are also summarized in \cite{schmaler_cresst_2010}.} For other materials, which are, however, not discussed in this work, we rely on the Helm form factor \cite{helm_inelastic_1956} in the parametrization of Lewin/Smith \cite{lewin_review_1996}.

The total DM signal results from the weighted sum over all target nuclides, where the weight for each nuclide is its mass fraction.

\subsection{Excess light events} \label{subsec:excesslight}

Excess light events are events with an additional light component which may e.g.~be created by a particle traversing the scintillating polymeric foil surrounding our detector modules (see figure \ref{fig:scheme}) and depositing energy in the foil and the target crystal. Here, we follow the empirical parametrization of \cite{schmaler_cresst_2010} modeling the density of the excess light events as exponentially decreasing with $El_\text{dec}$ and distance to the mean of the electron band ($El_\text{width}$). Convolving with the resolution then yields:

\begin{eqnarray}\label{eq:excesslight}
\rho_\text{excess}\left(E,L\right)  = El_\text{amp}\exp\left(-\frac{E}{El_\text{dec}}\right)  \\
\cdot  \frac{1}{2 El_\text{width}}\exp\left(-\frac{L}{El_\text{width}}+\frac{\left(\sigma_\text{L,e}\right)^2}{2 El_\text{width}^2}\right) \nonumber \\
\cdot\left[1+\erf\left(\frac{L}{\sqrt{2}\sigma_\text{L,e}}-\frac{\sigma_\text{L,e}}{\sqrt{2}El_\text{width}}\right)\right] \nonumber \\
\text{with } \sigma_\text{L,e} = \sigma_\text{L}\left(L_\text{e}\left(E\right)\right) \nonumber
\end{eqnarray}
$El_\text{amp}$ is the amplitude of the density.

\subsection{Exposure and efficiency} \label{sec:efficiency}

\begin{figure}
    \centering
    \includegraphics[width=\columnwidth]{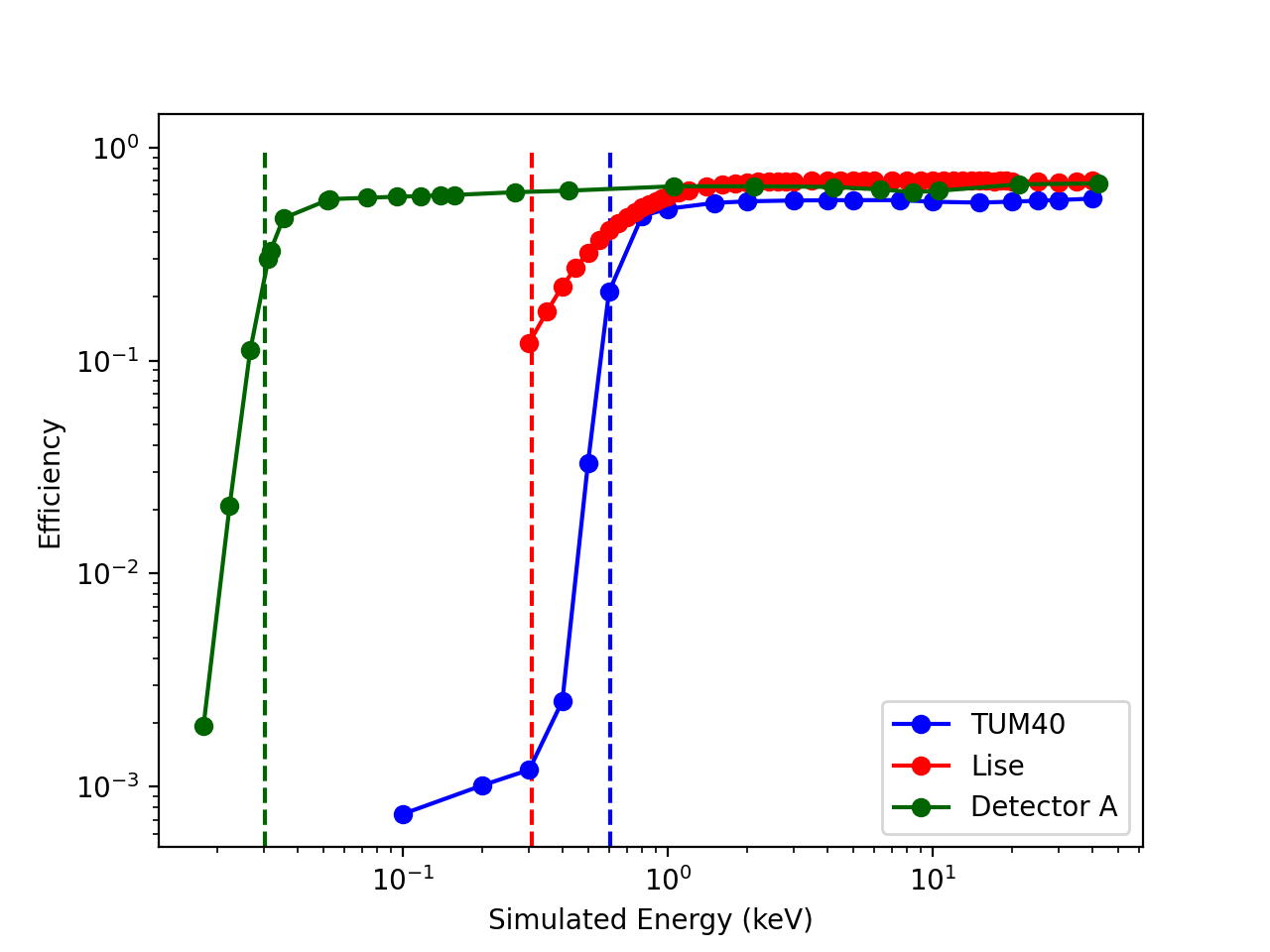}
    \caption{Efficiencies (= survival probabilities) for the detectors analyzed in this work as a function of injected (simulated) energy. The dashed vertical lines mark the energy threshold of the respective detector. The efficiencies were determined in the analyses in \cite{abdelhameed_first_2019,angloher_results_2016,reindl_exploring_2016}.}
    \label{fig:efficiencies}
\end{figure}

Table \ref{tab:detectorsanddata} lists the gross exposures of the respective data sets, which are given by the live time of the measurement times the mass of the target crystal. In the raw data analysis, we apply certain cuts to the data removing time periods of unstable detector operation, events where the correct determination of energy might not be guaranteed (e.g.~pile-up, tilted baselines, etc.) and events coincident between multiple detectors and/or the muon and/or iStick veto (see \cite{abdelhameed_first_2019,angloher_results_2016,reindl_exploring_2016}). The efficiency of this selection process, defined as the probability of a valid event to survive, is measured with artificially created pulses. These pulses are generated by scaling a pulse template to the desired injected energy and adding it to an empty baseline (a noise trace). Then the selection criteria are applied, and the fraction of surviving to totally generated events yields the efficiency $\text{eff}(E)$ for a certain energy $E$. The cuts are designed such that $\text{eff}(E)$ only depends on the energy. Its shape is the same for background data and neutron calibration data; the scaling of the latter is a free fit parameter, as will be discussed in the next subsection \ref{subsec:backgroundandneutroncalibrationdata}. Figure \ref{fig:efficiencies} shows the efficiencies as a function of energy in double-logarithmic scale.

Although we obtain and plot the efficiencies as a function of simulated (true) energy, we apply it to the reconstructed (measured) energy, that is after convolution with the resolution. The efficiencies as a function of simulated energy already account for the finite resolution of the detector, thus they have to be applied to the reconstructed energy. This is conservative and was e.g.~justified in \cite{abdelhameed_first_2019} with a more detailed study.

\subsection{Background and neutron calibration data} \label{subsec:backgroundandneutroncalibrationdata}

For all detectors analyzed in this work, we make use of the neutron calibration data, acquired by illuminating the experiment with an AmBe source. We always fit background data and neutron data simultaneously; the total negative log-likelihood is then given by: 

\begin{eqnarray} \label{eq:sumbckncal}
 %-\log\mathcal{L} = -\log \left[\sum_{i=0}^{i=N_\text{bck}} \text{eff}(E_i) \cdot \left( \sum_{j=0}^{j=N_{\rho_\text{bck}}}  \rho_j(E_i,L_i) \right) \right] + \\ \nonumber 
 -\log\mathcal{L} =  \sum_{i=0}^{i=N_\text{bck}} -\log\left[ \text{eff}(E_i) \cdot \left( \sum_{j=0}^{j=N_{\rho_\text{bck}}}  \rho_j(E_i,L_i) \right)  \right ] \\ \nonumber 
+ \sum_{k=0}^{k=N_\text{ncal}} -\log \left[ \text{eff}(E_k) \cdot \left( \sum_{l=0}^{l=N_{\rho_\text{ncal}}}  \rho_l(E_k,L_k) \right) \right] \\ \nonumber
+ \sum_{j=0}^{j=N_{\rho_\text{bck}}}   \iint_\text{acc. region} \text{d}E\, \text{d}L\, \text{eff}(E)\,  \rho_j(E,L) \\ \nonumber 
+ \sum_{l=0}^{l=N_{\rho_\text{ncal}}}  \iint_\text{acc. region} \text{d}E\, \text{d}L\, \text{eff}(E)\,  \rho_l(E,L)
\end{eqnarray}

with 

\begin{eqnarray} \label{eq:rho}
  \rho_x(E,L) = \frac{\text{d}N_x}{\text{d}E} \mathcal{N}\left(L_x\left(E\right),\sigma_x^2\left(E\right)\right)(L) \\
\mathcal{N}: \text{Gaussian function}  \nonumber
\end{eqnarray}

$N_\text{bck/ncal}$ is the number of events in the acceptance region for background and neutron calibration data, respectively. Accordingly, $N_{\rho_\text{bck/ncal}}$ is the number of densities to consider for the respective data.

The likelihood function in equation \ref{eq:sumbckncal} consists of four terms which we will briefly describe starting with the first one in the uppermost line. For each event, with energy $E_i$ and light $L_i$, we evaluate the densities $\rho_j(E_i,L_i)$ and sum the result which we multiply with the efficiency $\text{eff}(E_i)$. Then, for the total likelihood, one has to multiply these values for each event. Since we are using the $\log\mathcal{L}$, this multiplication simplifies to a sum of the logarithms. In the first term, we consider all events in the background data (index $i$) and all densities relevant to the background data (index $j$). The second term in line two does the same for the neutron calibration data (indices $k$ and $l$, respectively).

The last two terms are needed since the densities $\rho_{j,l}$ are not normalized, allowing one to have a different number of observed to expected events. The numbers of expected events are calculated in terms three and four for background and neutron calibration data, respectively. This formalism is known as extended maximum likelihood; see \cite{barlow_extended_1990} for further details. 

The acceptance region is the area in the light versus energy plane accessible to the likelihood framework. It starts at the energy threshold of the respective detector and ends at an a priori chosen upper energy, typically corresponding to the linear range of the detector (for the data analyzed in this work, see section \ref{sec:Results}). In light of this, we chose very loose bounds for the acceptance region, ensuring that all valid particle events are inside.

The density functions $\rho_\text{ncal}$ describing the neutron calibration data are constructed using the same band functions, consisting of mean lines and widths, as for the background density functions $\rho_\text{bck}$. Since all background contributions are also present during the neutron calibration, we add them to $\rho_\text{ncal}$ weighted by the relative exposure of neutron calibration to background data. However, in the neutron calibration, additional contributions may arise from the neutrons themselves, from secondary particles created by the neutrons or the neutron source, and contributions from a not fully closed shielding during the calibration.\footnote{To properly illuminate the detectors, the AmBe neutron source has to be placed inside the outer neutron veto in the inner region of the experimental setup. During this operation, the shieldings might not be completely closed, causing a slightly increased background level from external radiation. More details may be found in the Appendix \ref{app:fitresults}.}

Although external measurements of the quenching factors in \cawo~exist, only fitting the neutron calibration data for each detector allows us to determine the detector-specific factor $\epsilon$ precisely. For this reason, the use of the neutron calibration data in the likelihood fit is crucial to precisely determine the position of the nuclear recoil bands, which is critical to calculate the expected distribution of a potential DM signal. Another benefit of using neutron calibration data is the additional statistics (also in the electron and $\upgamma$-bands) which helps to fit the band parameters more accurately.

\subsection{Combination of detectors} \label{subsec:combination}

Similar to the combined description of background and neutron calibration data for a single detector, it is also possible to combine multiple detectors into a single likelihood description. In this case, the total log-likelihood is the sum of the log-likelihoods for the individual data sets. Unlike for the combination of background and neutron calibration data, the different detectors do not share the same band parameters. Also, the background and neutron calibration spectra are different. Instead, the only common parameters are present in the possible DM signal, and for detectors with the same material, the parameters describing the energy-dependent nuclear recoil quenching factors. Even though the properties of potential DM particles are the same, the shape and height of the dark matter spectrum can vary greatly between detectors due to different exposure, resolution, and efficiency.

The common dark matter signal also highlights the possible advantage of combining data sets from different detectors into a single description. The additional data provides more exposure which may result in stronger limits. For detectors with different target materials, the distinct shapes of the expected DM recoil spectra can yield a better signal-to-background separation, especially if a low-energy excess is present. Additionally, a combination of detectors could potentially help to take advantage of the different properties of multiple detectors, e.g.~two detectors, one with a low threshold and one with low background.

\subsection{Profile likelihood ratio} \label{subsec:PRL}

To set exclusion limits on the DM-nucleon cross-section, we are using the profile likelihood ratio (PLR) which is a standard statistical method of the field and which will be reported briefly here following the notation of \cite{cowan_asymptotic_2011}. 

The PLR is defined as: 
\begin{equation}
    \lambda(\mu) = \frac{\mathcal{L}(\mu,\hat{\hat{\bm{\theta}}})}{\mathcal{L}(\hat{\mu},\hat{\bm{\theta}})} 
    \label{eq:PLR}
\end{equation} 
where $\mu$ is the primary parameter of interest, in our case the DM-nucleus cross-section $\mu=\sigma_0$ (see equation \ref{eq:dmsignal}). The nuisance parameters are put to the vector $\bm{\theta}$.

In simple words, two (conditional) maximizations of the likelihood $\mathcal{L}$ are performed in the PLR. $\hat{\hat{\bm{\theta}}}$ is the set of nuisance parameters which maximizes the likelihood $\mathcal{L}$ for a given, fixed value of $\mu$. The denominator is the maximum likelihood for free $\hat{\mu}$ and $\hat{\bm{\theta}}$. 

The PLR can now be used to either \textit{claim a discovery} or set an exclusion limit. For discovery, one would use $\lambda(\mu=\sigma_0=0)$, so compare the null hypothesis (no DM signal) to the best fit (including a potential dark matter signal) using the test statistics
 \begin{equation}
     q_0 =
     \begin{cases}
       -2\ln\lambda(0) & \quad \text{if} \quad \hat{\mu} \ge 0\\
       0               & \quad \text{if} \quad \hat{\mu} <0\\
     \end{cases}
      \label{eq:TestStatisticsDiscovery}
 \end{equation}
 According to Wilk's theorem, $q_0$ follows a $\chi^2$ distribution with one degree of freedom \cite{wilks_large-sample_1938}. The statistical significance $Z$ is approximately given by $Z=\sqrt{q_0}$. 
 
 For limit setting the test statistics is very similar:
 
  \begin{equation}
   q_{\mu} =
     \begin{cases}
       -2\ln\lambda(\mu) & \quad \text{if} \quad \hat{\mu} \le \mu\\
       0                    & \quad \text{if} \quad \hat{\mu} > \mu\\
     \end{cases}
   \label{eq:TestStatLimit}
 \end{equation}
 
 To set a limit, we solve the above equation \ref{eq:TestStatLimit} to find the value for $\mu=\sigma_0$ corresponding to the desired and pre-defined confidence $Z$ (conventionally \unit[90]{\%} are used in direct DM detection).
 
\section{Results} \label{sec:Results}

In this paper, we apply the likelihood framework to three detectors; details on the detectors and data sets are given in section \ref{sec:CRESSTdetectors} and table \ref{tab:detectorsanddata}, the efficiencies are plotted in figure \ref{fig:efficiencies}. For reasons of clarity, we only show the fit results for TUM40 here. However, all fit results are available in the appendix \ref{sec:appendix} and the ancillary files. TUM40 has a low background level with clearly identifiable peaks and a well-performing light detector, but also a pronounced non-proportionality effect and a low-energy excess. These features make it the most challenging to fit out of the three detector modules considered in this work.

 The energy range for the acceptance region (see equation \ref{eq:sumbckncal})  always starts at the threshold. The upper energy boundary is chosen in accordance with previous work: \unit[16]{keV} for Detector A \cite{abdelhameed_first_2019} and \unit[40]{keV} for TUM40 \cite{angloher_results_2014}, and Lise \cite{angloher_results_2016}. For TUM40, we also give results (in the ancillary files) for a fit up to \unit[400]{keV} to demonstrate that our likelihood framework provides an accurate description of the data not only in the energy range relevant for dark matter but throughout the whole linear range of the detectors. The acceptance region boundaries in light are set generously to ensure that all relevant event contributions are fully contained in the acceptance region. The same acceptance region is set for background and neutron calibration data.

%Lise on the contrary has no noticeable low energy excess, barely any non-proportionality and an even lower threshold, but high background levels and bad light detector performance mean that a lot of events are in the signal region. Detector A of CRESST-III is currently the lowest threshold detector of CRESST and was used in a recent  publication \cite{firstresultscresstiii}, its distinct properties are similar to TUM40 just with an even stronger low energy excess, which dominates the background of this module. Details on the modules may also be found in table \ref{tab:detectorsanddata}.

%As a first step a maximum likelihood fit for all detector modules is performed and the result of this fit is plotted together with the data. Next, exclusion limits on the dark matter nucleon cross-section will be calculated for all detectors individually and compared against the optimum interval method \cite{yellin_finding_2002}. As a final step the benefits and disadvantages of combining detectors will be evaluated by comparing the combined limits to the individual ones.

\subsection{Band fits}

%\flo{Note: float placement in this section may not be ideal and will be optimized with the final text.}
For all band fits, various minimization methods, as well as multiple reruns, were used to ensure a global minimum (the most likely set of parameters) is found. All relevant parameters are free in the fit, except for the energy-dependent nuclear recoil quenching factors and the baseline resolutions for phonon ($\sigma_{\text{P},0}$) and light detector ($\sigma_{\text{L},0}$). While the likelihood value does not provide a way to judge the quality of the fit, a comparison of the resulting bands and spectra to the data can help evaluate the fit. 

\begin{figure}[t]
    \centering
    \includegraphics[width=\columnwidth]{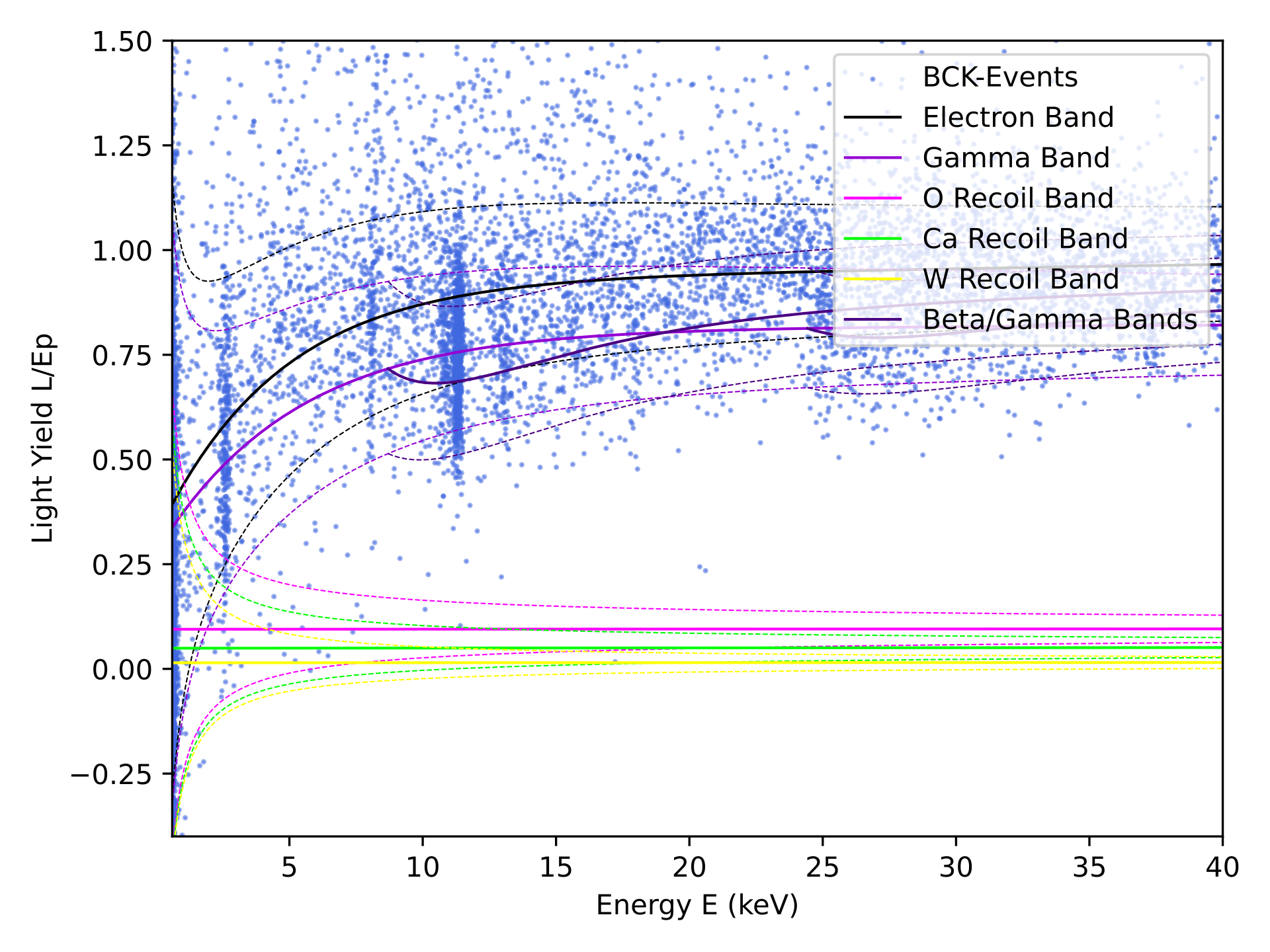}
    \caption{Result of the band-fit along with the \textbf{background data} of TUM40. The mean lines of the bands are shown as solid lines, while the dashed lines represent the lower and upper 90\% lines. The population of events above the electron band is mostly attributed to excess light events.}
    \label{fig:TUM40LYbck}
\end{figure}

\begin{figure}[t]
    \centering
    \includegraphics[width=\columnwidth]{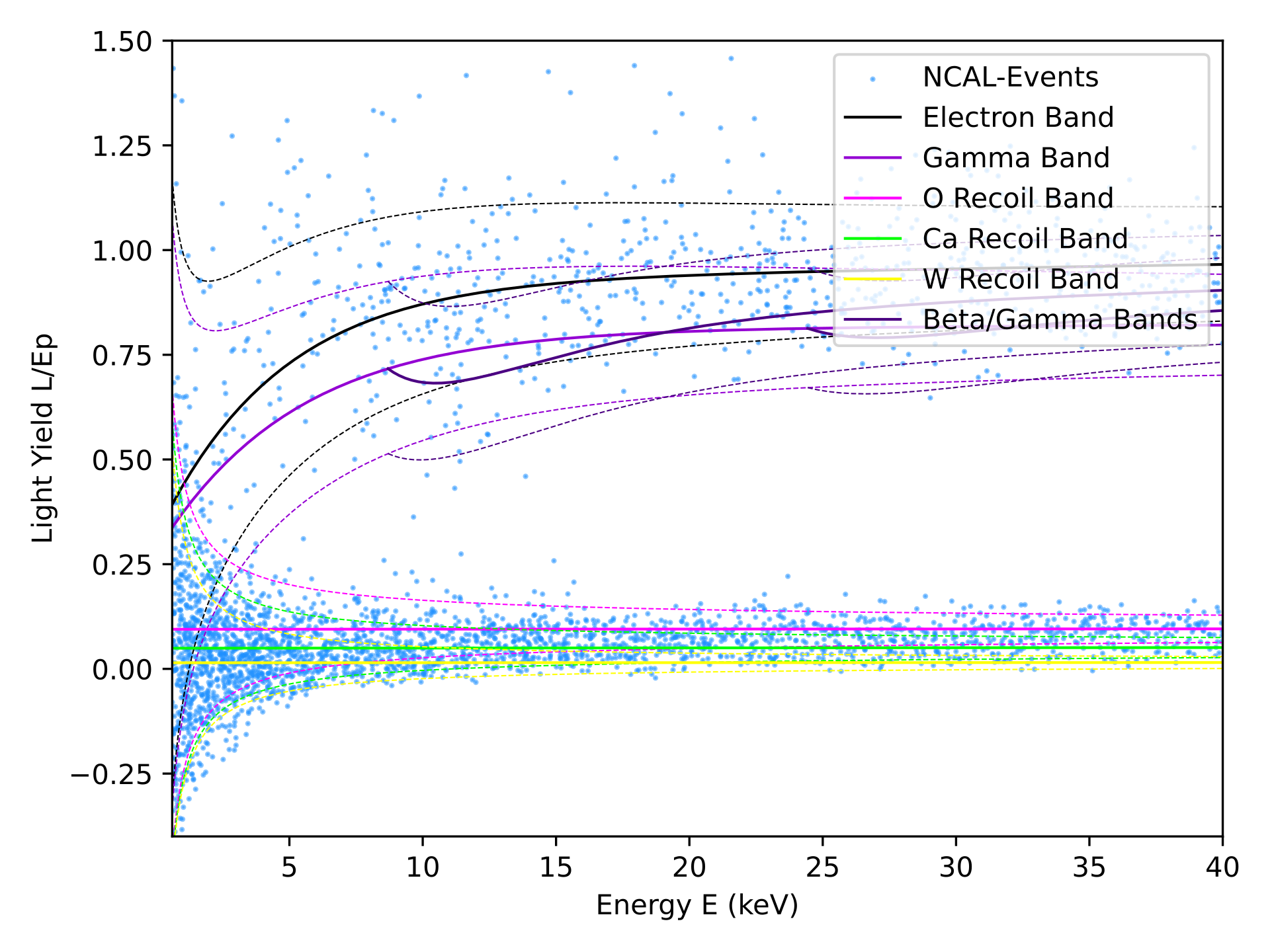}
    \caption{Result of the band-fit along with the \textbf{neutron calibration data} of TUM40. The mean lines of the bands are shown as solid lines, while the dashed lines represent the lower and upper 90\% lines.}
    \label{fig:TUM40LYncal}
\end{figure}

Figure \ref{fig:TUM40LYbck} shows the background data for TUM40 together with the resulting bands in the $LY$-$E$ plane. Figure \ref{fig:TUM40LYncal} shows the same bands but for the neutron calibration data. As one can see, the model accurately describes the position of the electron and the nuclear recoil bands. Also, the energy and light distribution of the $\upgamma$-lines (see section \ref{subsubsec:gammaband}), in particular at \unit[2.6]{keV} and \unit[11.3]{keV}, are very well described by the likelihood framework. It is also worth mentioning again that the fit can correctly determine the value of the scintillation efficiency $\eta$, see equation~\ref{eq:phononantiquenching}, resulting in straight, vertical $\upgamma$-lines particularly visible in figure \ref{fig:TUM40LYbck}.

\FloatBarrier %causes the floats to be flushed here

\begin{figure}[t!]
\includegraphics[width=1.0\linewidth]{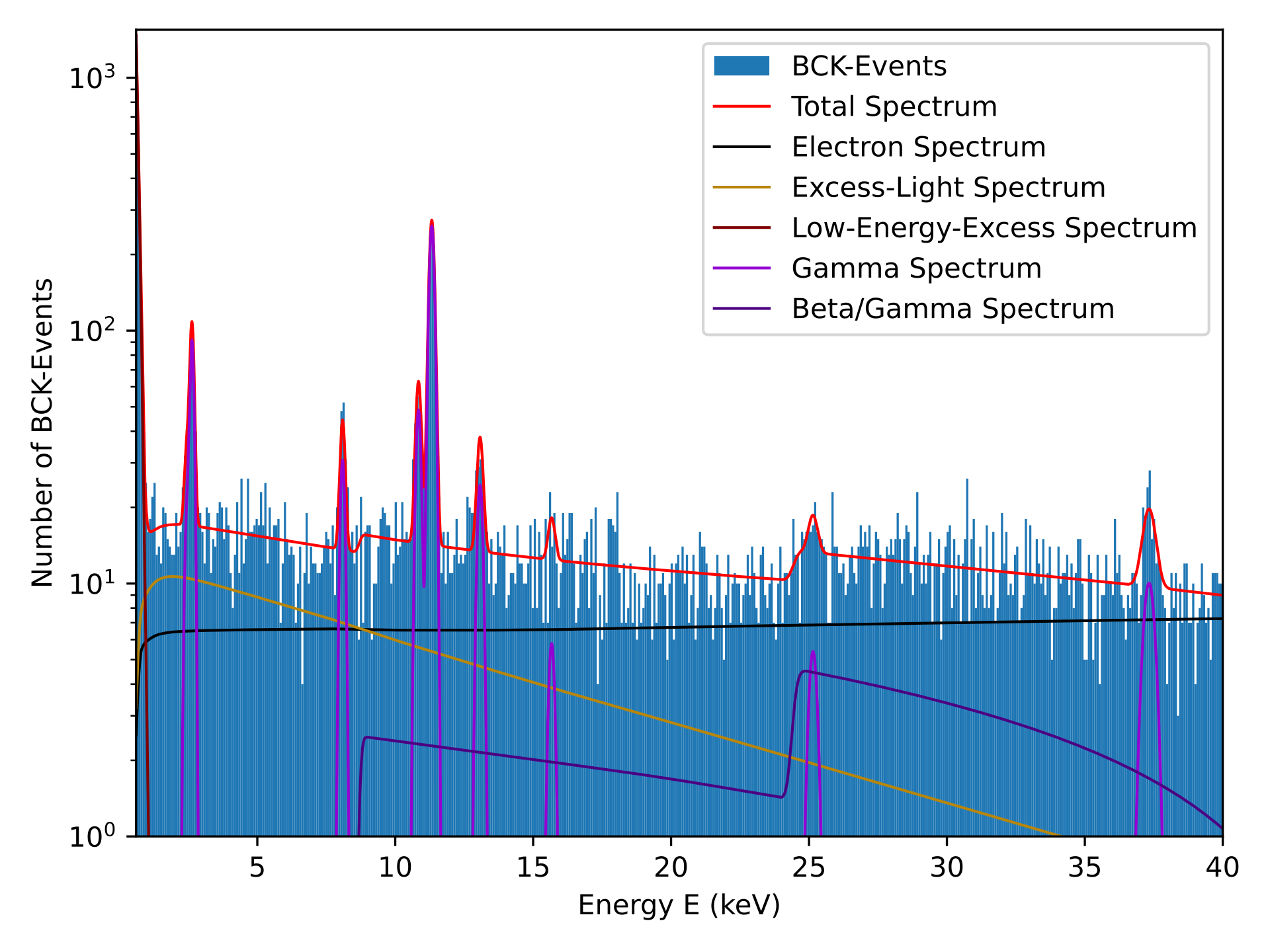}
\caption{Energy histogram of the \textbf{background data} of TUM40 including fit results.}
\label{fig:TUM40especbck}
\end{figure}

\begin{figure}[t!]
\includegraphics[width=1.0\linewidth]{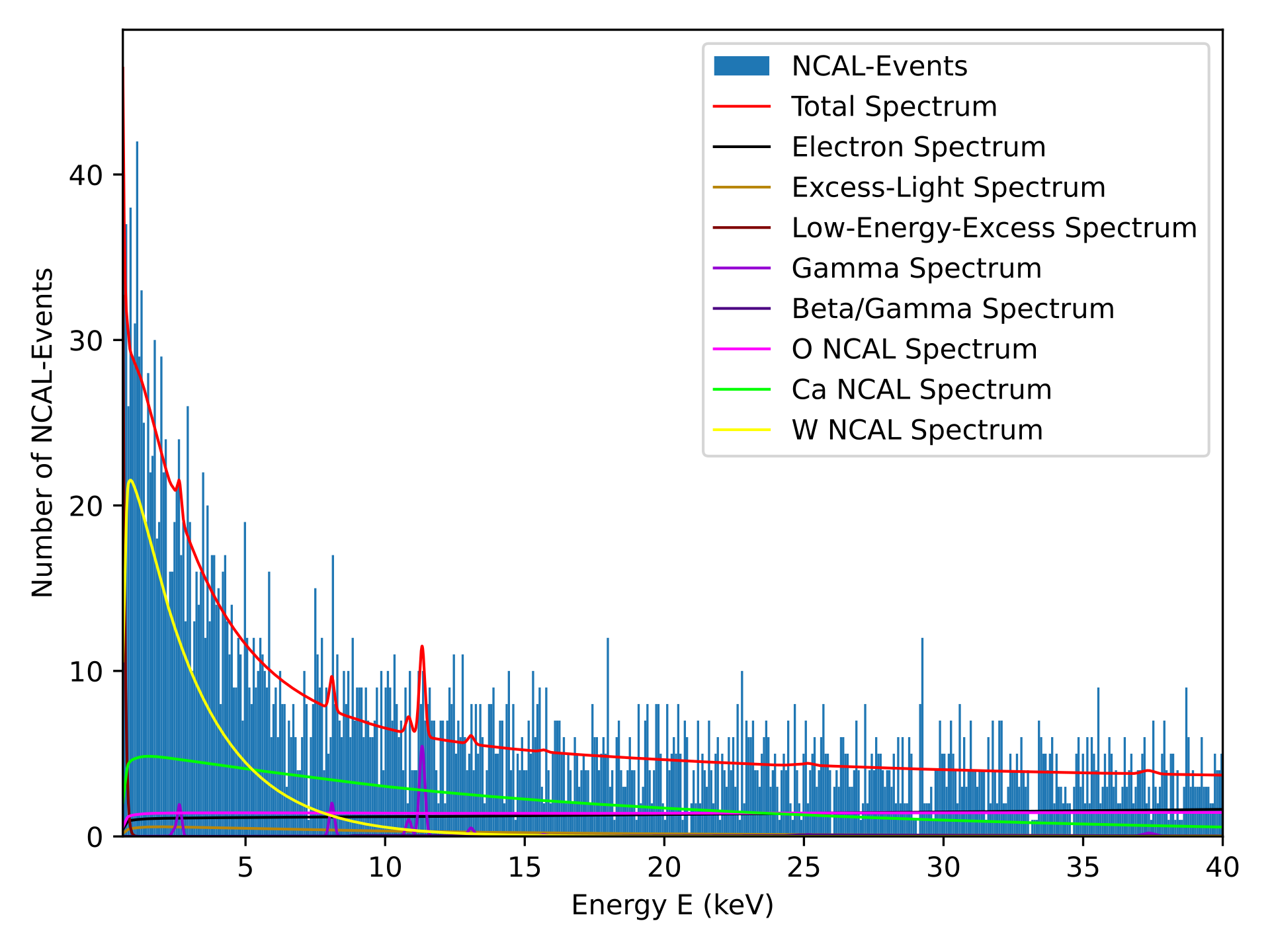}
\caption{Energy histogram of the \textbf{neutron calibation data} of TUM40 including fit results.}
\label{fig:TUM40especncal}
\end{figure}

Figures \ref{fig:TUM40especbck} and \ref{fig:TUM40especncal} show the energy histograms for the background and neutron calibration data, respectively, together with the fit results. Also these plots show that the likelihood framework is capable of accurately modeling the $\upgamma$-peaks, the $\upbeta$-decays, the ENR in the ncal data, and also the low energy excess. 
%\flo{TODO: Add information on the number of observed vs. predicted events.}

The resulting fit values, including uncertainties, are compiled in table \ref{tab:fitresults_TUM40}. Plots for the remaining detectors (Lise and Detector A) can be found in appendix \ref{sec:appendix}, CSV-files with the fit results are attached as ancillary files.

\subsection{Dark matter exclusion limits}

Exclusion limits at \unit[90]{\%} confidence level for the DM par\-ticle-nucleon cross-section are calculated using the profile likelihood method. All fits in this routine are performed using a global (particle swarm) and a local (Nelder-Mead) minimization. The same free parameters as for the band fit are used, except for high-energy $\upgamma$ and $\upbeta$/$\upgamma$ peaks. They do not overlap with the possible signal region and can therefore be fixed in the limit calculation to increase the convergence speed of the fits.

\begin{figure}
    \centering
    \includegraphics[width=1.0\linewidth]{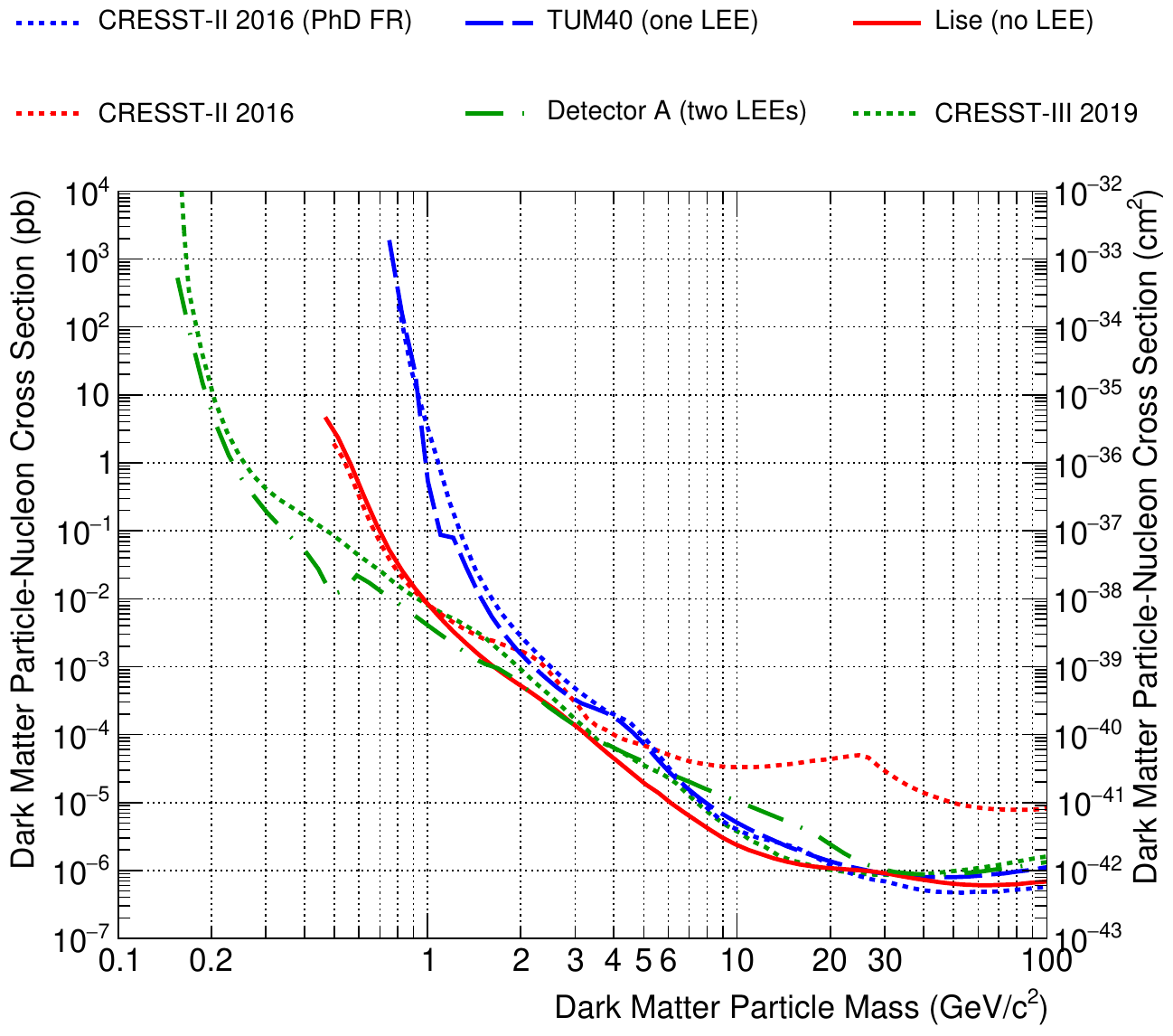}
    \caption{Exclusion limits for the detectors TUM40 (blue), Lise (red), and Detector A (green). Dotted lines are limits previously published by the CRESST collaboration for the same detectors (TUM40 \cite{reindl_exploring_2016}, Lise \cite{angloher_results_2016}, and Detector A   \cite{abdelhameed_first_2019}), but calculated with Yellin's optimum interval method \cite{yellin_finding_2002}. Likelihood limits for the respective detectors from this work are depicted as solid, dashed, and dashed-dotted lines. So, same-colored lines are based on the same data (from the same detector) but calculated with different methods: dashed for Yellin's optimum interval (previous works) and solid for likelihood limits (this work). 
}
    \label{fig:llvsyellin}
\end{figure}

To evaluate the performance of the profile likelihood, figure \ref{fig:llvsyellin} compares the likelihood limits to published limits which were obtained using the Yellin optimum interval method \cite{yellin_finding_2002}. As one can see, the likelihood and Yellin approaches generally lead to comparable limits for our data sets. One exception is the limit of Lise (red) which is considerably stronger in the likelihood case. This outcome is expected, as Lise had a rather poorly performing light detector which leads to a significant leakage from the \ega-bands to the nuclear recoil bands. The likelihood framework is able to model this background contribution, while the Yellin method treats it as a potential signal contribution weakening the resulting exclusion limit. It should be stressed that the likelihood limits for TUM40 and DetA only achieve comparable performance to the Yellin limits if (one or two, respectively) LEE excess contributions are allowed in the framework. Including the LEE is delicate, as its origin is not completely understood and its shape is similar (but not identical) to a potential DM signal. However, our knowledge of the LEE strongly disfavors a DM origin, justifying including an LEE background. In appendix \ref{sec:appendix}, the reader may find a more detailed discussion including an illustration of the influence of the LEE on the exclusion limits. 

The combination of detectors (see subsection \ref{subsec:combination}) in the likelihood framework is evaluated using the TUM40 and Lise data. The strong differences between the two detectors should help to identify the strengths and weaknesses of a combined calculation. For low DM masses, where only Lise contributes to the sensitivity, the combined limit is identical to the Lise limit. For the remaining mass range, the combined limit typically ranges between the limits of two single detectors. This might not seem intuitive at first glance, but is a direct result of the profile likelihood ratio test, see equation \ref{eq:PLR}. In summary, we show that the likelihood framework allows the combination of data from different detectors, but also that this does not automatically lead to a gain in sensitivity, in particular if data from detectors with very different features and performances are combined.

%\begin{figure}
%\includegraphics[width=1.0\linewidth]{figures/ll2021_all}
%\caption{TODO}
%\label{fig:limits_all}
%\end{figure}

\begin{figure}
\includegraphics[width=1.0\linewidth]{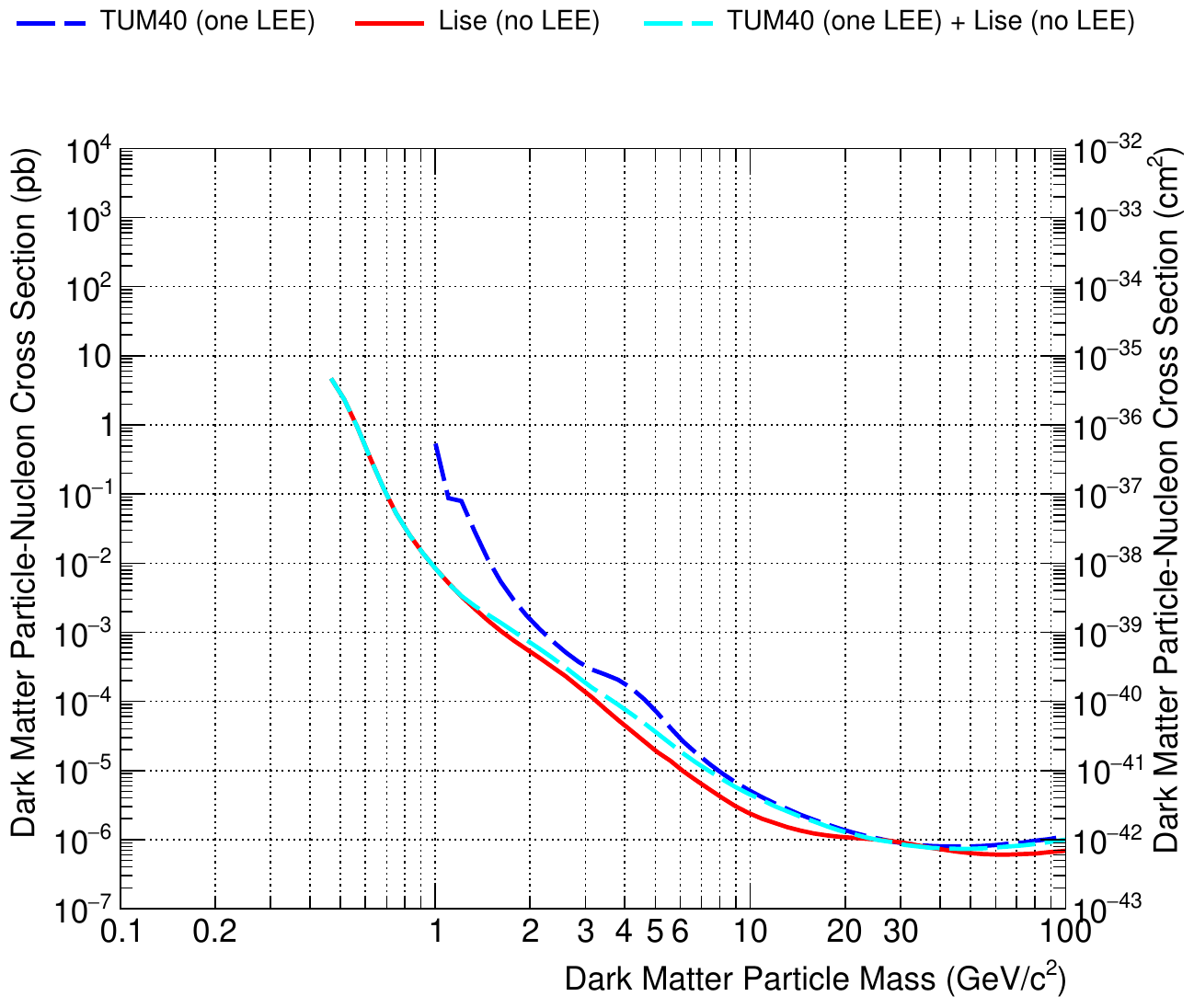}
\caption{Likelihood limits obtained for Lise (no LEE, red) and TUM40 (one LEE, dashed blue) and a limit obtained from the combined likelihood function (dashed cyan).}
\label{fig:limits_combined}
\end{figure}

%alternative version for combination of limits
%\begin{figure}
%\ theincludegraphics[width=1.0\linewidth]{figures/Combined01092022_full.pdf}
%\caption{Alternative to \ref{fig:limits_combined}, but with limits without LEE. Flo: I would prefer \ref{fig:limits_combined}.}
%\label{fig:limits_combined_alternative}
%\end{figure}

\section{Conclusion and outlook} \label{sec:Conclusion}

In this paper, we present a profile maximum likelihood framework for the description of data from scintillating cryogenic calorimeters in the scintillation light versus energy plane. We apply this model to data from three detectors from the CRESST DM experiment and find that the model accurately describes the data. The likelihood allows extracting precise information on the particle processes happening in the detector and the detector response, thereby making it a framework of very high value for the understanding and modeling of our detectors. 

Furthermore, it is the basis for the test of a potential DM signal in the data allowing for a discovery analysis as well as for setting limits on the DM-nucleon interaction cross-section. Since CRESST does not observe a potential DM signal in their data, we set exclusion limits in this paper and find comparable, or stronger, limits than those obtained previously using the Yellin optimum interval method. In addition, the likelihood framework enables the combination of data from different detectors in a straightforward way, which we also showcased here. 

The likelihood framework discussed in this paper may lay ground for many future developments, such as, in particular, combining data from many detectors ($\mathcal{O}(100)$) for the planned upgrade of the CRESST readout system. In addition, it may be used to test alternative DM models and to include background estimates obtained with Monte Carlo simulations directly \cite{abdelhameed_geant4-based_2019,angloher_high-dimensional_2023}. We also plan to extend the likelihood to include the time information as a third event observable to allow for properly evaluating time-dependent signals (e.g.~the DM modulation signature) and backgrounds (e.g.~radioactive decays, or the low-energy excess\cite{angloher_latest_2023}).

\section*{Acknowledgments}

This work has been funded by the Deutsche Forschungsgemeinschaft (DFG, German Research Foundation) under Germany’s Excellence Strategy – EXC 2094 – 390783311 and through the Sonderforschungsbereich (Collaborative Research Center) SFB1258 ‘Neutrinos and Dark Matter in Astro- and Particle Physics’, by the BMBF 05A20WO1 and \linebreak 05A20VTA, and by the Austrian science fund (FWF): I5420-N and P 33026-N {AnaCONDa}. FW was supported through the Austrian research promotion agency (FFG), project \linebreak ML4CPD. SG was supported through the FWF project \linebreak {STRONG-DM} (FG1). JB and HK were funded through the FWF project P 34778-N ELOISE. The Bratislava group acknowledges partial support provided by the Slovak Research and Development Agency (projects APVV-15-0576 and \linebreak {APVV-21-0377}). We are grateful to the Laboratori Nazionali del Gran Sasso - INFN for their generous support of CRESST. 

\section{Appendix} \label{sec:appendix}
\subsection{Numerical minimization tools}
 
The large number of parameters and their strong correlation present a significant challenge when maximizing the likelihood.

The efficiency is not an analytical function and therefore prevents the use of an analytical derivative. The nature of the log-likelihood function which is a sum over many data points makes a numerical estimation of the gradient neither accurate nor fast. Therefore, most of the algorithms used are gradient-free methods.
Additionally, to assure the validity of the determined parameters as well as the profile likelihood ratio, precise convergence to the global maximum is important. No single algorithm is capable of handling such a diverse and challenging use case while being sufficiently fast. For this reason, a multitude of algorithms is used in this work. Since most optimization problems require the finding of a minimum, optimization methods are usually implemented as a minimization. The negative log-likelihood is used as an objective function for the minimization algorithms.
The most important algorithms for this work and their implementation are presented in the following. A brief introduction to the method is presented; we refer to the corresponding reference for a more detailed explanation.

Two packages which implement various minimization algorithms included are used. The first one is the \texttt{Optim.jl} package \cite{mogensen2018optim}, which contains various algorithms mainly focused on unconstrained optimization of uni- and multivariate functions. In addition, we use \texttt{BlackBoxOptim.jl} \cite{bboptim2} and specializes in gradient-free, global optimization using meta-heuristic and stochastic algorithms.

\subsubsection{Differential evolution}

Differential Evolution optimization was first proposed by Storn and Price \cite{diffevl}. Similar to Particle Swarm optimization, Differential Evolution algorithms are population-based methods. The difference to Particle Swarm Optimization is the way in which candidate solutions are created. In Differential Evolution strategies, candidate solutions are created by adding the weighted difference vector of two candidate solutions to a third one.
The Differential Evolution algorithm used here is part of the "BlackBoxOptim.jl" \cite{bboptim2} package; the implementation is called "Adaptive DE/rand/1/bin with radius limited sampling".
This implementation does not take any starting values, instead, it probes the whole search space. Due to its fast and robust global convergence as well as not introducing any bias due to starting values, it is extremely well suited for an initial estimate of the parameters for a new dataset.

\subsubsection{Particle swarm}

This method uses a population of candidate solutions denoted "a particle" in this context. This population moves through the entire search space to find a global minimum. The velocity and direction of individual particles are influenced by the optimal solution found by the particle itself, by the best solution found by the nearest particles, as well as the best global solution.
The implementation used here is called Adaptive Particle Swarm optimization and was first proposed in \cite{zhan2009adaptive}. This implementation improves global convergence compared to classical Particle Swarm implementations. The algorithm implements four evolutionary states, one of which is called 'jumping out', and improves global convergence at the cost of speed by moving particles away from current minima in search of better ones.
The implemented Particle Swarm optimization takes starting values and is, therefore, well-suited if a rough estimate for the parameters exists. In addition, due to the global convergence properties, it is more robust against local convergence compared to the Nelder-Mead algorithm. While convergence for a small number of dimensions tends to be slower compared to the Nelder-Mead method, for high dimensions or strong correlations between parameters the Particle Swarm optimization converges faster.

\subsubsection{Nelder-Mead}
The Nelder-Mead method \cite{nelder1965simplex} -- often called downhill simplex method --  is a gradient-free direct search method. A simplex is constructed, containing information about the function value at different points. From this point, the algorithm performs one of four possible operations: reflection, expansion, contraction or shrinking. The purpose of this operation is to gradually replace points with high function values with points with lower values.
The behavior of this algorithm can be tuned by tweaking the parameters associated with the four operations. The implementation used here utilizes the adaptive parameters introduced by Gao and Han \cite{gao2012implementing} which provide better convergence characteristics in high dimensions.
The algorithm usually provides fast and precise convergence to a local minimum. In this application, it is, therefore, usually used as the final step of the minimization to refine the minimum found by the global optimization routines.

\subsubsection{Minuit} \label{subsec:minuit}

Since the \texttt{Minuit} package is well established in the particle physics community, for a detailed description the reader is referred to \cite{james_minuit:_1998}. We use \texttt{MIGRAD} as an alternative or in addition to Nelder-Mead for the minimization of the negative log-likelihood function. We also use \texttt{MINUIT's} built-in output of parameter uncertainties, where the symmetric uncertainties are based on the error/covariance matrix and the asymmetric error is based on \texttt{MINOS}. 

\subsubsection{Other algorithms}

Various other minimization algorithms are implemented, but in most cases, they are outperformed by the ones mentioned before. An implementation of a conjugate gradient descent method \cite{CGD} as well as a simulated annealing algorithm \cite{SIMANN} from the "Optim.jl" package can be used. From the "BlackBoxOptim.jl" package a few natural evolution-based algorithms \cite{wierstra2011natural} and a generating set search method \cite{KoLeTo03} can be employed.

\subsection{Implementations} \label{subsec:implementations}

The results of this work were obtained with a non-public software named \texttt{Romeo}. \texttt{Romeo} is programmed in the \texttt{Julia} language \cite{_julia_2022}. To cross-check \texttt{Romeo} a second program, na\-med \texttt{lxx}, was developed independently. \texttt{lxx} is based on C++ and uses \texttt{ROOT} (version 6.18, \cite{brun_root_1997}) for plotting and for the interface to the \texttt{MINUIT} minimizer \cite{james_minuit:_1998}. \texttt{lxx} and \texttt{Romeo} both are in the ballpark of \unit[10,000]{lines} of code and reach similar speed\footnote{\texttt{Romeo} is about twice as fast as lxx.}. Their results on the best fit $\mathcal{L}(\hat{\mu},\hat{\bm{\theta}})$ perfectly agree within numerical accuracy.

\subsection{Effect of the low energy excess (LEE)} \label{sec:lee_effect}

\begin{figure}
    \centering
    \includegraphics[width=1.0\linewidth]{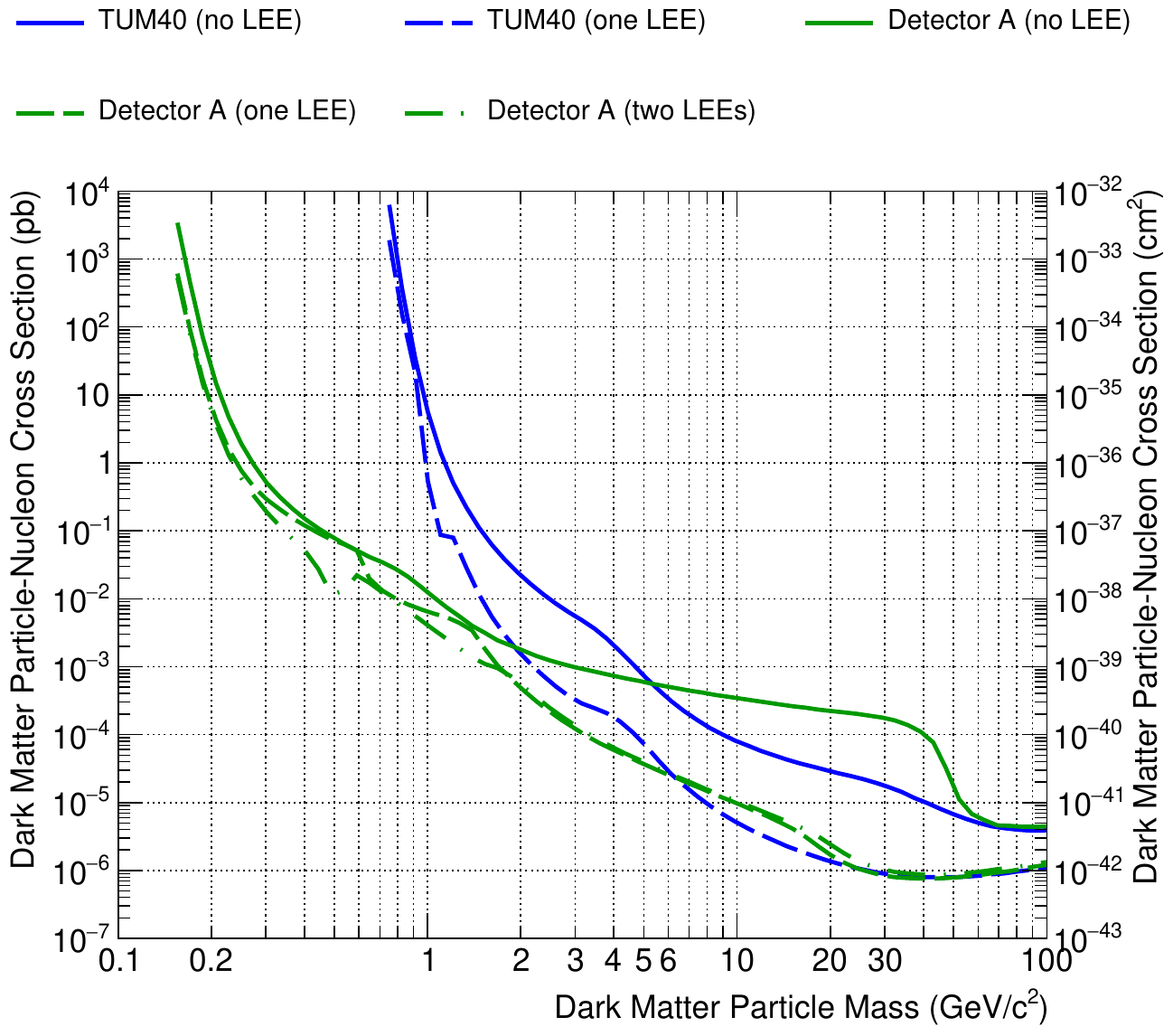}
    \caption{Likelihood limits for TUM40 and Detector A without low-energy-excess (LEE) in solid, with one LEE component in dashed and two LEE components (only for Detector A) in dashed-dotted. }
    \label{fig:limitsLEE}
\end{figure}

We observe an excess of events at low energies for the modules TUM40 and Detector A which was discussed in section \ref{subsubsec:lee}. In figure \ref{fig:limitsLEE} we show a comparison of limits obtained with likelihood frameworks with no LEE contribution (solid), one LEE contribution (dashed), and two LEE contributions (dashed-dotted, only Detector A). For both detectors, including one, or two LEE contribution(s) significantly improves the limit, in particular for DM particle masses larger than \unit[1]{\gev}. This is not surprising as the LEE is similar, but not identical, in shape to the DM spectrum. Therefore, if no LEE is allowed the likelihood fit is forced to increase the DM contribution to make model and data compatible, even if the shapes of the LEE and the DM spectrum are slightly different. 

It should be noted that the inclusion of a background similar in shape to the expected signal has to be done with special care. For TUM40, a DM origin of the LEE can be safely excluded, as Lise with an even lower threshold does not show any LEE. For Detector A, a DM origin can also be quite robustly rejected due to several observations, in particular, its rate decaying with time, as shown in \cite{angloher_latest_2023,adari_excess_2022}.

\section{Fit results} \label{app:fitresults}

Table \ref{tab:fitresults_TUM40} lists the results of the maximum likelihood fit (best fit) for the detector TUM40 with one LEE contribution for the energy range $[E_\text{thr}=\unit[0.6]{keV},\unit[40]{keV}]$.  The uncertainties were calculated with \texttt{MINUIT} (see section \ref{subsec:minuit}); the symmetric uncertainty is calculated via the covariance matrix and represents a symmetric uncertainty interval of the parameter around the best-fit value. The \texttt{MINOS} errors are also obtained with \texttt{MINUIT} and give an asymmetric uncertainty interval. The \texttt{MINOS} uncertainties are found by increasing/\-decreasing the parameter in question while simultaneously minimizing the negative log-likelihood for all other (N-1) parameters. The lower/upper uncertainty for a parameter then corresponds to an increase of the negative log-likelihood of 0.5 which in turn corresponds to one standard deviation. The main advantage of the computation-intensive \texttt{MINOS} procedure is that it fully takes into account correlations between parameters and non-linearities \cite{james_minuit:_1998}. \texttt{MINUIT} also provides a validity criterion (evaluated on maximal function calls, improvement on parameters and convergence) on the uncertainty interval which is listed in the last column. It should be noted that the scintillation efficiency $\eta$ (parameter 28 in table \ref{tab:fitresults_TUM40}) had to be fixed in the calculation of the \texttt{MINOS} errors to reach convergence. The value listed in the table, however, corresponds to the value obtained by the maximum likelihood fit. 

As already discussed, we fit the background data (bck) and the neutron calibration data (ncal) simultaneously. All background components have to be also present with the same or a higher rate in the ncal data. With the ratio of the exposures between ncal and bck (par.~26) one can turn the rate into an activity in the ncal data. This parameter is a free parameter in the fit, as it can be precisely determined by e.g. the cosmogenically activated $\upgamma$-lines. We assume the same energy dependence of the efficiency (see figure \ref{fig:efficiencies}) between bck and ncal data, but by leaving this parameter-free in the fit we can account for an overall lower efficiency in the ncal data which is expected due to the higher overall rate in the ncal data creating more pile-ups and impacting detector stability. The value obtained for par.~26 is however in very well agreement with plausibility checks done on the data. For some background processes, we find a higher rate in the ncal data, thus we add a second component of the respective process only increasing the rate, but not the position/shape and mark these parameters with "ncal" in table \ref{tab:fitresults_TUM40}. An example would be the Cu-fluorescence line at \unit[$\sim8$]{keV} (pars.~47-49) where the rate in ncal is roughly five times higher than the bck rate. This is expected, as the fluorescence is triggered by the overall background level. For background components with the same rate in bck and ncal we set the ncal contribution to zero. It should be stressed that the likelihood framework allows extracting the rates in bck and ncal directly from data which is very valuable information in the identification of the background components.
\clearpage

\onecolumn
{\small
%\begin{table*}
%    \centering

    \begin{filecontents*}{AncillaryFilesPublic/FitResults/TUM40/Detector_TUM40_PhD_Flo_Ep_Minos_parameters_converted.csv}
Name;Nameinpaper;Unit;Value;Fixed;Lower_Bound;Upper_Bound;Hesse_Error;Minos_Lower;Minos_Upper;Minos_Valid
np_decay;$L_{3}$;keV;4.380043264335479;False;0.0;50.0;0.5282539521662075;-0.4921866561848797;0.5312683267484647;True
np_fract;$L_{2}$;;0.6442828403267291;False;0.0;1.0;0.04233377470307975;-0.0389060728391209;0.04412300893414508;True
L0;$L_{0}$;keVeekeV$^{-1}$;0.9311433659172408;False;0.5;1.5;0.01854644107547382;-0.01787877906128299;0.01866541552234977;True
L1;$L_{1}$;keVeekeV$^{-2}$;0.000922912235610085;False;-0.01;0.01;0.0005517176642307895;-0.0005534836894242792;0.0005362500263471844;True
sigma_l0;$\sigma_{L,0}$;;0.246189;True;0.0;1000.0;0.00246189;;;
S1;$S_{1}$;keVee;0.2528557397360195;False;0.0;1.0;0.009674943017073251;-0.009619463098374257;0.00993021143962868;True
S2;$S_{2}$;;0.004048770160038678;False;0.0;0.01;0.0007633114253150927;-0.0007743174435362942;0.0007824182508276046;True
el_amp_bck;$El_{amp,bck}$;keV$^{-1}$;288.13750545010123;False;0.0;5000.0;14.504359118796144;-14.208285804353954;14.718838943550924;True
el_amp_ncal;$El_{amp,ncal}$;keV$^{-1}$;9.681683594407463;False;0.0;5000.0;2.4110029874903347;-2.4694544809975048;2.6297360998474804;True
el_decay;$El_{dec}$;keV;13.330654802748962;False;0.0;100.0;0.9040043435549984;-0.8495925040560647;0.9565404419359396;True
el_width;$El_{width}$;keVee;5.105986640966153;False;0.0;50.0;0.16674754032636985;-0.1637644535575562;0.1718463910890261;True
sigma_p0;$\sigma_{P,0}$;keV;0.0741;True;0.0;1000.0;0.000741;;;
sigma_p1;$\sigma_{P,1}$;;0.005657439558596872;False;0.0;0.05;0.0003740575029452368;-0.0003740949570744541;0.0003747045728697532;False
E_p0_bck;$P_{0,bck}$;keV$^{-1}$;145.4555048083897;False;0.0;5000.0;8.463375159499094;-8.534400569508552;8.633971051769386;True
E_p1_bck;$P_{1,bck}$;keV$^{-2}$;0.3634084894460301;False;-50.0;50.0;0.3479647062301652;-0.35489788112429377;0.3538088972801587;True
E_p0_ncal;$P_{0,ncal}$;keV$^{-1}$;21.169089455769786;False;0.0;5000.0;2.807783035745219;-2.9209315464221404;2.9995026640150777;True
E_p1_ncal;$P_{1,ncal}$;keV$^{-2}$;0.28550247715085675;False;-50.0;50.0;0.11282591040849255;-0.11603304671812968;0.11588282801822432;True
E_fr_bck;$A_{lee,bck}$;keV$^{-1}$;4688823297.282252;False;0.0;50000000000.0;2460031450.914618;-1269945410.5941565;2700244372.81427;False
E_fr_ncal;$A_{lee,ncal}$;keV$^{-1}$;5753205.957992337;False;0.0;50000000000.0;1018226.207318007;-5749506.6561635835;75744735.38942139;False
E_dc;$l_{lee}$;keV;0.0554746684713997;False;0.0;10.0;0.002414407373931282;-0.0026553115498451074;0.0014837289798569652;False
E_dc_ncal;$l_{lee,ncal}$;keV;-1.0;True;-2.0;10.0;-0.01;;;
L_lee;$L_{lee}$;keVeekeV$^{-1}$;0.2231286742311987;False;0.0;1.0;0.011459258018044538;-0.011470164904205882;0.011553338976003271;False
QF_y;$Q_{\gamma\,1}$;;0.8816234251601627;False;0.5;1.2;0.013018331146993078;-0.01312809189485948;0.013185716082683306;True
QF_ye;$Q_{\gamma,2}$;;-0.0004701351208379473;False;-0.01;0.01;0.0006934249434335975;-0.0006979707478416631;0.0007019172003398063;True
eps;$\epsilon$;;0.8557457791279454;False;0.5;1.0;0.01865290943248027;-0.01831508990265576;0.019357671226124067;True
ratio_ncal;$\frac{\text{Exp.(ncal)}}{\text{Exp.(bck)}}$;;0.02099797995781355;False;0.0;1.0;0.003920885876676633;-0.005633852712810722;0.00455505822355013;False
thr;$E_{thr}$;keV;0.6056;True;0.0;1.1;0.006056000000000001;;;
eta;$\eta$;keVkeVee$^{-1}$;0.06613806396875974;True;0.0;1.0;0.0006613806396875973;;;
n_O_fr_bck;$A_{O,bck}$;keV${-1}$;0.0;True;0.0;100.0;0.1;;;
n_O_dc_bck;$l_{O,bck}$;keV;32.45903519150707;True;0.0;200.0;0.3245903519150707;;;
n_O_fr_ncal;$A_{O,ncal}$;keV${-1}$;31.930927444820327;False;0.0;200.0;2.4093362373920666;-2.411965822906632;2.4093229416442217;False
n_O_dc;$l_{O}$;keV;999999.9977072786;False;0.0;1000000.0;521080.77251985593;-521035.6722546314;521035.6722546314;False
n_Ca_fr_bck;$A_{Ca,bck}$;keV${-1}$;0.0;True;0.0;250.0;0.1;;;
n_Ca_dc_bck;$l_{Ca,bck}$;keV;17.8814462130457;True;0.0;100.0;0.178814462130457;;;
n_Ca_fr_ncal;$A_{Ca,ncal}$;keV${-1}$;121.83475074844242;False;0.0;5000.0;16.49687998886359;-15.526589458643372;16.547600974064633;False
n_Ca_dc;$l_{Ca}$;keV;17.521667539312848;False;0.0;100.0;1.9626303406836223;-1.7625476293324809;2.145713899821801;True
n_W_fr_bck;$A_{W,bck}$;keV${-1}$;0.0;True;0.0;1000.0;0.1;;;
n_W_dc_bck;$l_{W,bck}$;keV;0.0;True;0.0;1000.0;0.1;;;
n_W_fr_ncal;$A_{W,ncal}$;keV${-1}$;788.6282142109546;False;0.0;20000.0;70.30043027343778;-66.44540402464801;74.35728730564745;True
n_W_dc;$l_{W}$;keV;2.4208598942028456;False;0.0;10.0;0.22321920897376304;-0.2164907199830298;0.223816797517882;True
FG_1_C_bck;$C_{\gamma,1,bck}$;keV${-1}$;73.68963861629874;False;0.0;1000.0;22.278366029852396;-22.67408884043018;22.67408884043018;False
FG_1_C_ncal;$C_{\gamma,1,ncal}$;keV${-1}$;0.0;True;0.0;100.0;0.1;;;
FG_1_M;$\mu_1$;keV;2.437365989823169;False;2.3;2.7;0.03170605672208615;-0.031861600864072505;0.03392253644805576;True
FG_2_C_bck;$C_{\gamma,2,bck}$;keV${-1}$;388.6519474357718;False;0.0;500.0;30.618615532041222;-34.14665284678525;33.33766863178049;True
FG_2_C_ncal;$C_{\gamma,2,ncal}$;keV${-1}$;1.5445419465631712e-05;False;0.0;100.0;12.8522697942825;-9.080864074584304e-05;5.412037267381514;True
FG_2_M;$\mu_2$;keV;2.6291079015954697;False;2.5;2.9;0.00808074884741683;-0.007993258029269378;0.008671874971897364;True
FG_3_C_bck;$C_{\gamma,3,bck}$;keV${-1}$;150.5488902119869;False;0.0;500.0;20.075884573930963;-20.100088644790432;20.100088644790432;False
FG_3_C_ncal;$C_{\gamma,3,ncal}$;keV${-1}$;6.7233633332769775;False;0.0;100.0;5.24563401602415;-5.927927514456486;6.651225365905017;False
FG_3_M;$\mu_3$;keV;8.090475211111277;False;7.8;8.3;0.012102317791459429;-0.012114788496456003;0.012122981223198137;True
FG_4_C_bck;$C_{\gamma,4,bck}$;keV${-1}$;267.75709048064306;False;0.0;500.0;26.10115418465897;-26.10937221771381;26.10937221771381;False
FG_4_C_ncal;$C_{\gamma,4,ncal}$;keV${-1}$;0.0;True;0.0;100.0;0.1;;;
FG_4_M;$\mu_4$;keV;10.842495275129625;False;10.5;11.1;0.012281167215219213;-0.012281564409973631;0.012303611332733113;False
FG_5_C_bck;$C_{\gamma,5,bck}$;keV${-1}$;1457.5964486863706;False;0.0;3000.0;53.40652499816826;-52.94602320309875;54.216966684058164;True
FG_5_C_ncal;$C_{\gamma,5,ncal}$;keV${-1}$;0.0;True;0.0;200.0;0.1;;;
FG_5_M;$\mu_5$;keV;11.324459360214089;False;11.0;11.6;0.003800373197475615;-0.003805103552774813;0.003798524955855179;True
FG_6_C_bck;$C_{\gamma,6,bck}$;keV${-1}$;147.87607953748122;False;0.0;500.0;20.97088012628931;-20.335764424025776;20.9765793641226;False
FG_6_C_ncal;$C_{\gamma,6,ncal}$;keV${-1}$;0.0;True;0.0;100.0;0.1;;;
FG_6_M;$\mu_6$;keV;13.074419553213122;False;12.7;13.4;0.020845435363956;-0.021138785798260833;0.02066663300050017;True
FG_7_C_bck;$C_{\gamma,7,bck}$;keV${-1}$;38.592574045921204;False;0.0;500.0;14.337821979798285;-13.63329392384381;15.087300590637287;True
FG_7_C_ncal;$C_{\gamma,7,ncal}$;keV${-1}$;0.0;True;0.0;100.0;0.1;;;
FG_7_M;$\mu_7$;keV;15.669091495572145;False;15.2;16.0;0.05818184285283135;-0.06030442346174246;0.058175168566621416;False
FG_8_C_bck;$C_{\gamma,8,bck}$;keV${-1}$;48.827534688238856;False;0.0;500.0;18.184921128928845;-17.528331442580804;18.99971685214623;True
FG_8_C_ncal;$C_{\gamma,8,ncal}$;keV${-1}$;0.0;True;0.0;100.0;0.1;;;
FG_8_M;$\mu_7$;keV;25.14421860786657;False;24.8;25.6;0.07834705813377951;-0.08406678854487112;0.07756318217774659;True
FG_9_C_bck;$C_{\gamma,9,bck}$;keV${-1}$;124.39975823859994;False;0.0;500.0;22.200723274772397;-21.448360414186435;23.11338003923627;True
FG_9_C_ncal;$C_{\gamma,9,ncal}$;keV${-1}$;0.0;True;0.0;100.0;0.1;;;
FG_9_M;$\mu_7$;keV;37.328054028099295;False;36.0;38.0;0.03791786016418541;-0.03804410965442792;0.03803098853857904;True
FB_1_C;$C_{\beta/\gamma,1}$;;738.2616936654683;False;0.0;2000.0;89.42404209221905;-88.5411111994832;89.66104274819168;False
FB_1_M;$E_{\gamma,1}$;keV;24.435545091058497;False;24.0;25.0;0.12164851345444914;-0.1237457643520336;0.12495674973018912;True
FB_1_D;$D_{1}$;keV;20.3;True;19.5;23.0;0.203;;;
FB_2_C;$C_{\beta/\gamma,2}$;;999.7672975025173;False;0.0;2000.0;129.13007200207932;-129.2659252279345;128.7663158204757;False
FB_2_M;$E_{\gamma,2}$;keV;8.698120773779442;False;8.5;10.0;0.07904335517932283;-0.16552801024568986;0.12493085913915075;True
FB_2_D;$D_{2}$;keV;35.5;True;35.0;38.0;0.355;;;
\end{filecontents*}
\pgfplotstabletypeset[
    create on use/Number/.style = {create col/expr accum={\pgfmathaccuma+1}{0}},
    begin table=\begin{longtable},
    end table=\end{longtable},
    %every  head  row/.append  style={after  row={\caption{The  caption}},
    every last row/.append style={after row={\caption{Maximum likelihood fit results for TUM40 including uncertainties. The column "Fixed" indicates whether a parameter was free in the fit (Fixed = False), and the lower and upper boundaries mark the fit boundaries for the parameters. The parameter $\eta$ was free for the minimization but had to be fixed for the MINOS error calculation. A false MINOS validity indicates a potential issue in the calculation of the asymmetric MINOS uncertainty. In case, MINOS could not find a minimum at all, it returns the symm.~uncertainty, otherwise, it returns its best estimate. For more details, the reader is referred to \cite{noauthor_root_nodate-1}.}\label{tab:fitresults_TUM40}}},
    col sep=semicolon, 
   % unbounded coords=jump,
    header = has colnames,
    forcemathmode/.style={%
                preproc cell content/.append style={/pgfplots/table/@cell content/.add={$}{$}},},
    columns={Nameinpaper,Unit,Value,Fixed,Lower_Bound,Upper_Bound,Hesse_Error,Minos_Lower,Minos_Upper,Minos_Valid},
    %columns/Number/.style={column name ={}},
   % columns/Name/.style={string type,forcemathmode},
   % columns/Name/.style={string type,column type ={l}},
    columns/Nameinpaper/.style={column name ={\textbf{Name}}, string type, column type ={l}},
    columns/Unit/.style={column name = {\textbf{Unit}}, string type, column type ={l}},
    columns/Value/.style={column name = {\textbf{Value}}, dec sep align, precision = 2, std},
    columns/Fixed/.style={column name ={\textbf{Fixed}}, string type, column type ={l}, 
    string replace*={True}{{\color{blue}True}},
    string replace*={False}{{\color{Orchid}False}}
    },
    columns/Lower_Bound/.style={column name ={\textbf{Lower bound}}, dec sep align, precision = 2, std },
    columns/Upper_Bound/.style={column name ={\textbf{Upper bound}}, dec sep align, precision = 2, std },
    columns/Hesse_Error/.style={column name ={\textbf{Symm. unc.}}, dec sep align, precision = 2, std},
    columns/Minos_Lower/.style={column name ={\textbf{\texttt{MINOS} low}}, dec sep align, precision = 2, std},
    columns/Minos_Upper/.style={column name ={\textbf{\texttt{MINOS} upper}}, dec sep align, precision = 2, std},
    columns/Minos_Valid/.style={column name ={\textbf{\texttt{MINOS} validity}}, string type, column type ={l}, 
    string replace*={True}{{\color{ForestGreen}True}}, string replace*={False}{{\color{red}False}}},
    ]{AncillaryFilesPublic/FitResults/TUM40/Detector_TUM40_PhD_Flo_Ep_Minos_parameters_converted.csv}
%\end{table*}

}

%\bibliographystyle{h-physrev}

%\bibliographystyle{apsrev4-1}
%\bibliographystyle{h-physrev}

%\bibliography{bibflo,bibdaniel}
\printbibliography

%\clearpage
%
%\input{comments_internal_review.tex}
\end{document}

%% file: authors_cresst_EPJ_format.tex
\author{
  G.~Angloher\thanksref{addrMPI}\and
  S.~Banik\thanksref{addrHEPHY,addrAI}\and
  G.~Benato\thanksref{addrLNGS,addrGSSI}\and
  A.~Bento\thanksref{addrMPI,addrCoimbra}\and 
  A.~Bertolini\thanksref{addrMPI}\and 
  R.~Breier\thanksref{addrBratislava}\and
  C.~Bucci\thanksref{addrLNGS}\and 
  J.~Burkhart\thanksref{addrHEPHY}\and
  L.~Canonica\thanksref{addrMPI,addrBicocca}\and 
  A.~D'Addabbo\thanksref{addrLNGS}\and
  S.~Di~Lorenzo\thanksref{addrMPI}\and
  L.~Einfalt\thanksref{addrHEPHY,addrAI}\and
  A.~Erb\thanksref{addrTUM,addrWMI}\and
  F.~v.~Feilitzsch\thanksref{addrTUM}\and 
  S.~Fichtinger\thanksref{addrHEPHY}\and
  D.~Fuchs\thanksref{addrMPI}\and 
  A.~Garai\thanksref{addrMPI}\and 
  V.M.~Ghete\thanksref{addrHEPHY}\and
  P.~Gorla\thanksref{addrLNGS}\and
  P.V.~Guillaumon\thanksref{addrMPI,addrSaoP}\and
  S.~Gupta\thanksref{addrHEPHY}\and 
  D.~Hauff\thanksref{addrMPI}\and 
  M.~Ješkovsk\'y\thanksref{addrBratislava}\and
  J.~Jochum\thanksref{addrTUE}\and
  M.~Kaznacheeva\thanksref{addrTUM}\and
  A.~Kinast\thanksref{addrTUM}\and
  H.~Kluck\thanksref{addrHEPHY}\and
  H.~Kraus\thanksref{addrOxford}\and 
  S.~Kuckuk\thanksref{addrTUE}\and
  A.~Langenk\"amper\thanksref{addrMPI}\and 
  M.~Mancuso\thanksref{addrMPI}\and
  L.~Marini\thanksref{addrLNGS}\and 
  B.~Mauri\thanksref{addrMPI}\and
  L.~Meyer\thanksref{addrTUE}\and
  V.~Mokina\thanksref{addrHEPHY}\and
  M.~Olmi\thanksref{addrLNGS}\and
  T.~Ortmann\thanksref{addrTUM}\and
  C.~Pagliarone\thanksref{addrLNGS,addrCASS}\and
  L.~Pattavina\thanksref{addrLNGS,addrBicocca}\and
  F.~Petricca\thanksref{addrMPI}\and 
  W.~Potzel\thanksref{addrTUM}\and 
  P.~Povinec\thanksref{addrBratislava}\and
  F.~Pr\"obst\thanksref{addrMPI}\and
  F.~Pucci\thanksref{addrMPI, addrTUM}\and 
  F.~Reindl\thanksref{ef,addrHEPHY,addrAI} \and
  J.~Rothe\thanksref{addrTUM}\and 
  K.~Sch\"affner\thanksref{addrMPI}\and 
  J.~Schieck\thanksref{addrHEPHY,addrAI}\and 
  D.~Schmiedmayer\thanksref{ed,addrHEPHY,addrAI}\and 
  S.~Sch\"onert\thanksref{addrTUM}\and 
  C.~Schwertner\thanksref{addrHEPHY,addrAI}\and
  M.~Stahlberg\thanksref{addrMPI}\and 
  L.~Stodolsky\thanksref{addrMPI}\and 
  C.~Strandhagen\thanksref{addrTUE}\and
  R.~Strauss\thanksref{addrTUM}\and
  I.~Usherov\thanksref{addrTUE}\and
  F.~Wagner\thanksref{addrHEPHY,addrETH}\and 
  V.~Wagner\thanksref{addrTUM}\and
  V.~Zema\thanksref{addrMPI}
(CRESST Collaboration)
}

\institute
{Max-Planck-Institut f\"ur Physik, D-85748 Garching, Germany \label{addrMPI} \and
Institut f\"ur Hochenergiephysik der \"Osterreichischen Akademie der Wissenschaften, A-1050 Wien, Austria\label{addrHEPHY} \and
Atominstitut, Technische Universit\"at Wien, A-1020 Wien, Austria \label{addrAI} \and
INFN, Laboratori Nazionali del Gran Sasso, I-67100 Assergi, Italy \label{addrLNGS} \and
Comenius University, Faculty of Mathematics, Physics and Informatics, 84248 Bratislava, Slovakia \label{addrBratislava} \and
Physik-Department, TUM School of Natural Sciences, Technische Universit\"at M\"unchen, D-85747 Garching, Germany
\label{addrTUM} \and
Eberhard-Karls-Universit\"at T\"ubingen, D-72076 T\"ubingen, Germany \label{addrTUE} \and
Department of Physics, University of Oxford, Oxford OX1 3RH, United Kingdom \label{addrOxford} \and
also at: LIBPhys-UC, Departamento de Fisica, Universidade de Coimbra, P3004 516 Coimbra, Portugal \label{addrCoimbra} \and
also at: Walther-Mei\ss ner-Institut f\"ur Tieftemperaturforschung, D-85748 Garching, Germany \label{addrWMI} \and
also at: GSSI-Gran Sasso Science Institute, I-67100 L'Aquila, Italy \label{addrGSSI} \and
also at: Dipartimento di Ingegneria Civile e Meccanica, Universit\`a degli Studi di Cassino e del Lazio Meridionale, I-03043 Cassino, Italy\label{addrCASS} \and
also at: Dipartimento di Fisica, Università di Milano Bicocca, Milano, I-20126, Italy \label{addrBicocca} \and
also at: Instituto de Física, Universidade de São Paulo, São Paulo 05508-090, Brazil\label{addrSaoP} \and
present address: Department of Physics, ETH Zurich, CH-8093 Zurich, Switzerland and ETH Zurich - PSI Quantum Computing Hub, Paul Scherrer Institute, CH-5232 Villigen, Switzerland \label{addrETH}}
\thankstext{ef}{corresponding author: florian.reindl@tuwien.ac.at}
\thankstext{ed}{corresponding author: daniel.schmiedmayer@tuwien.ac.at}